\documentclass[a4paper,11pt]{article}
\pdfoutput=1 

\usepackage{jheppub} 

\usepackage{amscd}
\usepackage{amsfonts}
\usepackage{amsmath}
\usepackage{amssymb}
\usepackage{amsthm}
\usepackage{accents}
\usepackage{youngtab}
\usepackage{braket}
\usepackage{titlesec}
\usepackage{cancel}
\usepackage[svgnames]{xcolor}
\usepackage{ytableau}
\usepackage{here}
\usepackage{tikz}
 \usetikzlibrary{patterns,arrows,calc,tikzmark}
\usepackage{feynmf}
\usepackage{mathtools}
\usepackage{pb-diagram}
\usepackage{simplewick}

\usepackage[labelformat=parens,subrefformat=parens]{subcaption}
\usepackage{comment}
\usepackage{ascmac} 

\DeclareMathOperator*{\Res}{Res}

\newcommand{\ba}{\begin{eqnarray}}
\newcommand{\ea}{\end{eqnarray}}

\titleclass{\subsubsubsection}{straight}[\subsection]

\newcounter{subsubsubsection}[subsubsection]
\renewcommand\thesubsubsubsection{\thesubsubsection.\arabic{subsubsubsection}}

\titleformat{\subsubsubsection}
  {\normalfont\normalsize\bfseries}{\thesubsubsubsection}{1em}{}
\titlespacing*{\subsubsubsection}
{0pt}{3.25ex plus 1ex minus .2ex}{1.5ex plus .2ex}

\makeatletter
\renewcommand\paragraph{\@startsection{paragraph}{5}{\z@}%
  {3.25ex \@plus1ex \@minus.2ex}%
  {-1em}%
  {\normalfont\normalsize\bfseries}}
\renewcommand\subparagraph{\@startsection{subparagraph}{6}{\parindent}%
  {3.25ex \@plus1ex \@minus .2ex}%
  {-1em}%
  {\normalfont\normalsize\bfseries}}
\def\toclevel@subsubsubsection{4}
\def\toclevel@paragraph{5}
\def\toclevel@paragraph{6}
\def\l@subsubsubsection{\@dottedtocline{4}{7em}{4em}}
\def\l@paragraph{\@dottedtocline{5}{10em}{5em}}
\def\l@subparagraph{\@dottedtocline{6}{14em}{6em}}
\makeatother

\makeatletter
\@addtoreset{equation}{subsection}
\makeatother

\graphicspath{{./figures/}}

\title{A Note on Quiver Quantum Toroidal Algebra}

\author[a]{Go Noshita,}
\author[a]{Akimi Watanabe}

\affiliation[a]{Department of Physics, The University of Tokyo,\\ 7-3-1 Hongo, Bunkyo-ku, Tokyo 113-0033, Japan}

\emailAdd{noshita@hep-th.phys.s.u-tokyo.ac.jp}
\emailAdd{awatanabe@hep-th.phys.s.u-tokyo.ac.jp}

\abstract{Recently, Li and Yamazaki proposed a new class of infinite-dimensional algebras, quiver Yangian, which generalizes the affine Yangian $\mathfrak{gl}_{1}$.  The characteristic feature of the algebra is the action on BPS states for non-compact toric Calabi-Yau threefolds, which are in one-to-one correspondence with the crystal melting models. These algebras can be bootstrapped from the action on the crystals and have various truncations. 
	
In this paper, we propose a $q$-deformed version of the quiver Yangian, referred to as the quiver quantum toroidal algebra (QQTA).  We examine some of the consistency conditions of the algebra. In particular, we show that QQTA is a Hopf superalgebra with a formal super coproduct, like known quantum toroidal algebras. QQTA contains an extra central charge $C$. When it is trivial ($C=1$),  QQTA has a representation acting on the three-dimensional crystals, like Li-Yamazaki's quiver Yangian. While we focus on the toric Calabi-Yau threefolds without compact 4-cycles, our analysis can likely be generalized to all toric Calabi-Yau threefolds.}

\begin{document} 
\maketitle

\section{Introduction}                                          
The study of the non-perturbative dynamics of supersymmetric gauge theories and superstring theory has been one of the most central topics in quantum field theories and string theory. Direct microscopic studies on four-dimensional supersymmetric gauge theories became accessible since the work of \cite{Nekrasov:2002qd}.
Rich algebraic structures lie behind the correspondences of four-dimensional supersymmetric gauge theories and other mathematical objects. AGT correspondence \cite{Alday2010} is one of them, and infinite-dimensional algebras such as Virasoro or W algebras' play became relevant. In particular, the affine Yangian of $\mathfrak{gl}_1$, which is shown to contain these conformal symmetries in a universal form, played an essential role in proving the equivalence \cite{schiffmann2012cherednik,Nakajima_Heisenberg,maulik2018quantum}.  

The $q$-deformed version of Virasoro/W algebra has been actively studied in mathematical literatures \cite{Awata:1996dx,Awata:1995zk,Feigin:1995sf,Shiraishi:1995rp}.
It was applied to the five-dimensional version of AGT correspondence\footnote{For other developments of this direction, see also \cite{Awata:2011ce,Kimura:2015rgi,Bourgine:2017jsi,Bourgine_2020,Bourgine_2016,Bourgine_2017,Awata_2016,Awata_2017RTT,Awata_2017,Awata_2018,harada2020quantum}.}\cite{awata2010five,Awata_2010,Yanagida_2010,awata2011notes}.
The universal symmetry which contains these $q$-deformed conformal symmetries is the quantum toroidal $\mathfrak{gl}_{1}$, which is sometimes referred to as ``DIM"  \cite{Ding:1996mq,Miki2007,Feigin2011,feigin2011quantum,Feigin_2012},
which is the $q$-deformation of the affine Yangian $\mathfrak{gl}_1$.
Compared with the undeformed case, the quantum toroidal $\mathfrak{gl}_1$ is more symmetric. Namely, it contains $SL(2,\mathbb{Z})$ duality, and also the Hopf algebra structure in a more natural form. Various extensions of DIM were done by mathematicians,
for instance, quantum toroidal $\mathfrak{gl}_{n}$ and $\mathfrak{gl}_{m|n}$ \cite{feigin2013representations,bezerra2019quantum}.

Recently, significant progress has been made on the relationship between infinite-dimensional algebras and Calabi-Yau geometry \cite{Rap_k_2019,rapcak2020cohomological,Li:2020rij,Galakhov:2020vyb}. For the case of the $\mathbb{C}^{3}$-geometry, we can define an action of the affine Yangian $\mathfrak{gl}_{1}$ \cite{schiffmann2012cherednik,Tsymbaliuk:2014,Prochazka:2015deb,Feigin_2012} on the BPS states.\footnote{See \cite{Gaiotto:2017euk,Prochazka:2017qum, Prochazka:2018tlo,Prochazka:2014gqa,Prochazka:2015deb} for corner vertex operator algebras and their relation with affine Yangian $\mathfrak{gl}_{1}$ and $W_{1+\infty}$. For gluing of affine Yangian $\mathfrak{gl}_{1}$, see also \cite{Gaberdiel_2018,Gaberdiel_2018_twin,Li_2020}.} Generalizations to BPS crystal configurations of general toric Calabi-Yau geometries are possible and the corresponding algebra is quiver Yangian \cite{Li:2020rij,Galakhov:2020vyb}.

The goal of this paper is to define the quiver quantum toroidal algebra (QQTA), which is the $q$-deformed version of \cite{Li:2020rij,Galakhov:2020vyb}  and study a class of its representations. We use the bootstrap method of \cite{Li:2020rij}, where the algebra is obtained through the action on three-dimensional BPS crystal \cite{Ooguri_2009}. It is a generalization of the MacMahon representation \cite{Feigin_2012} of quantum toroidal $\mathfrak{gl}_{1}$ \cite{Ding:1996mq, Miki2007, feigin2011quantum}.  While most of the analysis is parallel to \cite{Li:2020rij}, the $q$-deformed algebra has an extra structure such as an additional central charge and the coproduct structure, which should be directly compared with the mathematical literature \cite{feigin2013representations,bezerra2019quantum}. We also examined some of the consistency conditions of QQTA.  While we focus on the toric Calabi-Yau threefolds without compact 4-cycles, our analysis can likely be generalized to all toric Calabi-Yau threefolds.

This paper is organized as follows. In section \ref{sec:QYreview}, we review the properties of the quiver Yangian defined in \cite{Li:2020rij}. In section \ref{sec:quantum_toroidalgl1}, we review the properties of the well known quantum toroidal $\mathfrak{gl}_{1}$. It has a natural three-dimensional crystal representation, the MacMahon representation, and is a Hopf algebra. It becomes the affine Yangian $\mathfrak{gl}_{1}$, which is a quiver Yangian of $\mathbb{C}^{3}$ geometry in the degenerate limit.  In section \ref{sec:Algebra}, we define the quiver quantum toroidal algebra and see their properties. The definition given includes a central element $C$, but for now, this is a conjecture. It will be shown that even if we include the central element, it will be a Hopf superalgebra, which is similar to the quantum toroidal $\mathfrak{gl}_{1}$. In section \ref{sec:bootstrap}, we focus on the three-dimensional crystal and bootstrap the algebra when $C=1$ from it following \cite{Li:2020rij}. We also discuss generalizations when there are compact 4-cycles. In section \ref{sec:Example}, we introduce one new example. It is a quantum toroidal algebra associated with the orbifold $\mathbb{C}^{3}/(\mathbb{Z}_{2}\times\mathbb{Z}_{2})$. The quiver diagram of this algebra is the same as the Dynkin diagram of the affine superalgebra $D(2,1;\alpha)$. The natural three-dimensional crystal representation of this algebra is the same as the plane partition representation, but the pattern of the colors is only different. In section \ref{sec:summary}, we give a summary and some discussions for future work. The appendix is dedicated to basic facts of three-dimensional crystal melting and the convention we used in this paper. Some defining relations of QQTA are also written in detail.

\section{Review: Quiver Yangian}\label{sec:QYreview}
In \cite{Li:2020rij}, Li and Yamazaki constructed a new class of algebras, called quiver Yangian, from the toric diagram of the Calabi-Yau manifold. They associated the geometric data with the quiver diagram, through which they defined the algebra.
This section reviews the algebraic aspects of the quiver Yangian, which are directly relevant to the $q$-deformation.  To make this paper self-contained, we summarize the relation with the geometrical data in Appendix \ref{sec:3d_crystal}.
The quiver Yangian generalizes known affine Yangians, which includes the affine Yangian of $\mathfrak{gl}_1$ \cite{Tsymbaliuk:2014, schiffmann2012cherednik}, which played a significant role in proving 4D/2D duality.

\subsection{The definition of the quiver data}\label{sec:quiver-data}
We define the quiver Yangian from a set of data $Q=(Q_0, Q_1, Q_2)$, which is given by the toric diagram (see Appendix \ref{sec:3d_crystal}).
$Q_0$ is a set of vertices, and $Q_1$ is a set of arrows between the vertices. $Q_0$ and $Q_1$ define a quiver diagram. The set $Q_2$ consists of the loops constructed from the arrows in $Q_1$.\footnote{The periodic quiver diagram is a diagram with vertices, arrows, and faces, which are drawn on top of the torus. We denote $Q_{2}$ as the faces of this periodic quiver diagram. Each of the faces is a region surrounded by the arrows of $Q_{1}$, and we are identifying it with the sequence of arrows surrounding it  (see the Appendix \ref{sec:3d_crystal}).} It is associated with the superpotential in the context of string theory. We denote the number of elements of $Q_i$ ($i=1,2,3$) as $|Q_{i}|$, and they satisfy a relation
\begin{equation}
	|Q_0|-|Q_1|+|Q_2|=0,\label{eq:Eulerchar}
\end{equation}
which comes from the fact that the Euler number of $\mathbb{T}^{2}$ is 0.

\begin{figure}[h]
    \begin{tabular}{ccc}
      \begin{minipage}{0.3\hsize}
        \centering
        \includegraphics[width=4cm]{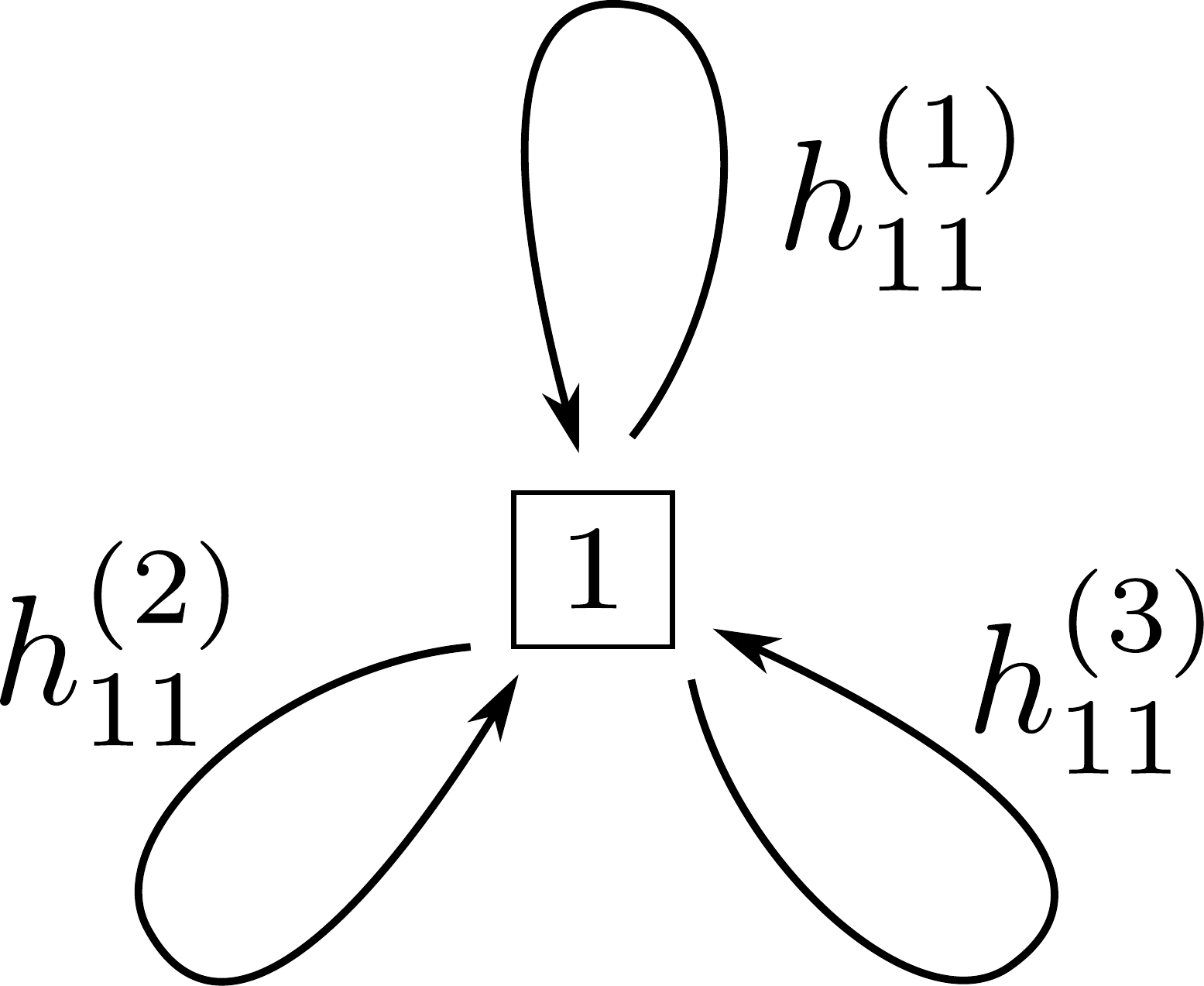}
        \subcaption{One vertex case}
      \end{minipage} &
      \begin{minipage}{0.3\hsize}
        \centering
        \includegraphics[width=4cm]{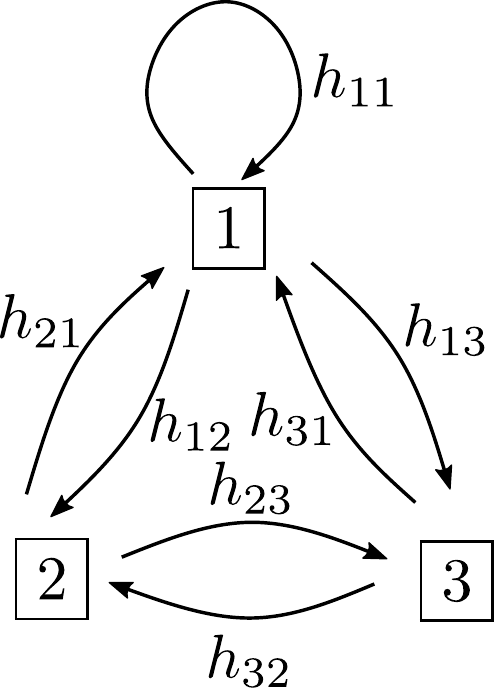}
        \subcaption{Three vertices and one self-loop case}
      \end{minipage} &
      \begin{minipage}{0.3\hsize}
        \centering
        \includegraphics[width=4cm]{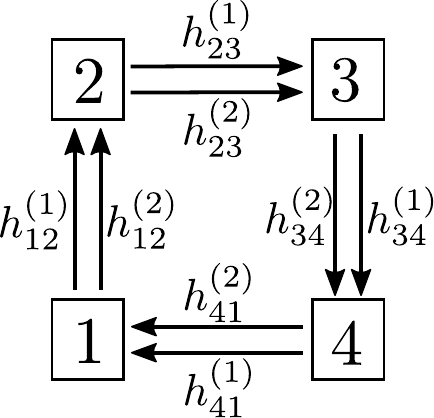}
        \subcaption{The case with $|i\to j|\neq |j\to i|$}
      \end{minipage}
    \end{tabular}
\caption{Three examples of the quiver diagrams. They have $|Q_0|$ vertices and $|Q_1|$ arrows.}
\label{fig:quiver_gl}
\end{figure}

Figure \ref{fig:quiver_gl} shows three examples of quiver diagrams.
For the first one (a), we have
\begin{equation}\label{eq:Q012_gl1}
	Q_0=\{1\},\quad Q_1=\{1\xrightarrow{1} 1,1\xrightarrow{2} 1,1\xrightarrow{3} 1\},\quad
	Q_2=\{1\xrightarrow{1} 1 \xrightarrow{2} 1 \xrightarrow{3} 1,1\xrightarrow{3} 1 \xrightarrow{2} 1 \xrightarrow{1} 1\}
\end{equation}
(see Figure \ref{fig:gl1Q2} of Appendix \ref{sec:3d_crystal}).
We note that there may be some arrows whose two ends are identical. We denote the number of arrows $i\to j$ in $Q_1$ as $|i\to j|$. In this example, $|1\to 1|=3$. We distinguish the arrows by adding an extra index over the arrow. We denote the parameter associated with the arrow $i\xrightarrow{a} j$ by $h^{(a)}_{ij}$, while we abbreviate the upper index when $|i\to j|=1$.

For the second one (b), 
\begin{align}
    Q_0&=\{1,2,3\},\quad
    Q_1=\{1\to 1, \; 1\to 2,\; 1\to 3,\; 2\to 1,\; 2\to 3,\; 3\to 1, \;3\to 2\},\\
    Q_2&=\{1\to1\to3\to1, 1\to2\to3\to2\to1, 1\to1\to2\to1, 1\to3\to2\to3\to1\}.
\end{align}
For the third one (c),
\begin{align}
	Q_0&=\{1,2,3,4\},\quad
	Q_1=\{1\xrightarrow{1}2,1\xrightarrow{2}2,2\xrightarrow{1}3,2\xrightarrow{2}3,
	3\xrightarrow{1}4,3\xrightarrow{2}4,4\xrightarrow{1}1,4\xrightarrow{2}1\},\\
	Q_2&=\{1\xrightarrow{a}2\xrightarrow{b}3\xrightarrow{c}4\xrightarrow{d}1\},\ \mbox{with }(a,b,c,d)\in \{(1,1,1,1),(2,2,2,2),(1,2,1,2),(2,1,2,1)\}.
\end{align}

We note that in the first two examples (a) and (b), $|i\to j|=|j\to i|$ for any $i,j\in Q_0$. In the third example (c), $|i\to j|\neq |j\to i|$ for some  pairs of elements in $Q_0$. We will refer to the quiver of the first (resp. second) type as ``symmetric" (resp. ``asymmetric").  In \cite{Li:2020rij}, the asymmetric quiver set comes from Calabi-Yau manifolds with compact four cycles.

We call the vertex $i\in Q_0$ as bosonic (resp. fermionic), when
$|i\to i|$ is odd (resp. even). In the quiver (b), two vertices are fermionic while one is bosonic. In the quiver (c), all the four vertices are fermionic.

\subsection{Deformation parameters and constraints}\label{sec:deformation-parameters}
For each element in $I\in Q_1$, we assign a deformation parameter $h_I$, which appears in the definition of quiver Yangian. There are constraints coming from each element in $Q_0$ (vertex constraint) and $Q_2$ (loop constraint).

\paragraph{Vertex constraints:}
\begin{equation}
	\sum_{I\in Q_1(a)}\mathrm{sign}_a(I) h_I =0,\label{eq:QYvertexconstr}
\end{equation}
where $a$ is an arbitrary vertex in $Q_0$.
$Q_1(a)$ implies the subset of $Q_1$ where the vertex $a\in Q_0$ is contained either in the start point or the endpoint. 
The $\mathrm{sign}_a(I)$ is $+1$ when $a$ is the endpoint of $I$, $-1$ when it is the start point, and $0$ when $I$ is the loop from $a$ to $a$.
For example, the vertex $1$ in the middle of Figure \ref{fig:quiver_gl} gives a constraint,
\begin{equation}
	h_{12}-h_{21}+h_{13}-h_{31}=0.
\end{equation}
Since the summation of the vertex constraints for all $a\in Q_0$ vanishes trivially, there are $|Q_0|-1$ independent constraints.

\paragraph{Loop constraints:}
\begin{equation}
	\sum_{I\in L} h_I=0,\label{eq:QYloopconstr}
\end{equation}
where $L$ is an arbitrary loop in the periodic quiver diagram. It is enough to impose the loop conditions on loops of $Q_{2}$ because arbitrary loops can be decomposed into loops of $Q_{2}$. Thus, we can consider $L$ as an arbitrary loop on $Q_{2}$ from now on. However, not all of the loop conditions (\ref{eq:QYloopconstr}) for loops in $Q_{2}$ are independent. 
For example, the loop constraints for the quiver set Figure \ref{fig:quiver_gl}(b) are:
\begin{align}
	h_{11}+h_{13}+h_{31}&=0,\label{eq:loopconst_eq1}\\
	h_{12}+h_{23}+h_{32}+h_{21}&=0,\label{eq:loopconst_eq2}\\
	h_{11}+h_{12}+h_{21}&=0,\label{eq:loopconst_eq3}\\
	h_{13}+h_{32}+h_{23}+h_{31}&=0.\label{eq:loopconst_eq4}
\end{align}
As the vertex constraints, a linear combination of the loop constraints becomes trivial. In the above example, (\ref{eq:loopconst_eq1})$+$(\ref{eq:loopconst_eq2})$-$(\ref{eq:loopconst_eq3})$-$(\ref{eq:loopconst_eq4})$=0$. Thus, there are $|Q_2|-1$ independent constraints.

To summarize, using (\ref{eq:Eulerchar}), the number of independent parameters becomes $|Q_1|-(|Q_0|-1)-(|Q_2|-1)=2$ for the quiver Yangian.

\subsection{The definition of quiver Yangian}
The Drinfeld currents of quiver Yangian are defined as
\begin{align}
    e^{(a)}(z) = \sum_{n=0}^\infty \frac{e_n^{(a)}}{z^{n+1}},\quad
    \psi^{(a)}(z) = \sum_{n=-\infty}^\infty \frac{\psi_n^{(a)}}{z^{n+1}},\quad f^{(a)}(z) = \sum_{n=0}^\infty \frac{f_n^{(a)}}{z^{n+1}},
\end{align}
with a formal expansion parameter $z\in \mathbb{C}$
for each $a\in Q_0$. The Drinfeld current with index $a$ is bosonic (resp. fermionic) if $a$ in $Q_0$ is the bosonic (resp. fermionic) vertex.

We define the bond factors as
\begin{equation}
    \varphi^{a\Rightarrow b}(u) \equiv \frac{\prod_{I\in\{b\to a\}} (u+h_I)} {\prod_{I\in\{a\to b\}} (u-h_I)},\label{eq:QYbondfactor}
\end{equation}
where $a, b\in Q_0$.
Set $\{a\to b\}$ is a subset of $Q_1$ which consists of the arrows from $a$ to $b$.
Using them, the OPE relations of the quiver Yangian are,
\begin{align}
\begin{split}
    \psi^{(a)}(z) \psi^{(b)}(w) &= \psi^{(b)}(w)\psi^{(a)}(z),\\
    \psi^{(a)}(z) e^{(b)}(w) &\simeq \varphi^{b\Rightarrow a}(z-w) e^{(b)}(w) \psi^{(a)}(z),\\
    e^{(a)}(z) e^{(b)}(w) &\sim (-1)^{|a||b|}\varphi^{b\Rightarrow a}(z-w) e^{(b)}(w) e^{(a)}(z),\\
    \psi^{(a)}(z) f^{(b)}(w) &\simeq \varphi^{b\Rightarrow a}(z-w)^{-1} f^{(b)}(w) \psi^{(a)}(z),\\
    f^{(a)}(z) f^{(b)}(w) &\sim (-1)^{|a||b|} \varphi^{b\Rightarrow a}(z-w)^{-1} f^{(b)}(w) f^{(a)}(z),\\
    [ e^{(a)}(z), f^{(b)}(w)\} &\sim -\delta^{a, b}\frac{\psi^{(a)}(z)-\psi^{(b)}(w)}{z-w},
    \end{split}
\end{align}
for $a, b\in Q_0$.
In the above equations, $\simeq$ means the equality up to $z^n w^{m\geq 0}$ terms and $\sim$ means the equality up to $z^{n\geq 0}w^m$ and $z^n w^{m\geq 0}$ terms. We note that the algebra is expected to be equipped with Serre relations, but for the moment, they are not known yet except for special cases. We will not touch on this topic in this paper. 

When both $a$ and $b$ are fermionic,
\begin{equation}
    [e^{(a)}(z), f^{(b)}(w)\} =e^{(a)}(z) f^{(b)}(w)+f^{(b)}(w) e^{(a)}(z).
\end{equation}
Otherwise,
\begin{align}
    [e^{(a)}(z), f^{(b)}(w)\} =e^{(a)}(z) f^{(b)}(w)-f^{(b)}(w) e^{(a)}(z).
\end{align}

One may define the quiver Yangian for both symmetric and asymmetric quiver sets. For the symmetric case, the bond factor becomes a homogeneous rational function, and the coefficients of $\psi^{(a)}(z)$ are simplified as
\begin{align}
    \psi_{n\leq -2}^{(a)}=0,\quad \psi_{-1}^{(a)}=1, \label{eq:nocpt4cycle}
\end{align}
which gives a usual expansion of $\psi$ generators in the affine Yangian algebras.

The quiver set for Figure \ref{fig:quiver_gl}(a) gives the simplest example of the quiver Yangian. It has one bosonic vertex and the loop constraint gives a constraint,
\begin{equation}
h^{(1)}_{11}+h^{(2)}_{11}+h^{(3)}_{11}=0\,,
\end{equation}
which is equivalent to the affine Yangian of $\mathfrak{gl}_{1}$, if we identify $h^{(a)}_{11}$ with the Nekrasov parameters.

The quiver set for Figure \ref{fig:quiver_gl}(b) has one bosonic and two fermionic vertices.  The quiver Yangian for this set gives the affine Yangian of  $\mathfrak{gl}_{2|1}$. One may generalize the quiver set to describe $\mathfrak{gl}_{m|n}$.  On the other hand, the data set such as Figure \ref{fig:quiver_gl}(c) gives a new family of algebras that may not be related to the Lie superalgebras.

\section{Review: Quantum toroidal \texorpdfstring{$\mathfrak{gl}_1$}{gl1}}\label{sec:quantum_toroidalgl1}
This article aims to study the $q$-deformations of the quiver Yangian. As mentioned in the previous section, the quiver Yangian contains the affine Yangian of $\mathfrak{gl}_{1}$ as a particular example. It plays a prototype of $q$-deformation, which motivate us to review the basic properties of the quantum toroidal $\mathfrak{gl}_{1}$, (sometimes it is referred to also as quantum continuous $\mathfrak{gl}_\infty$ or Ding-Iohara-Miki algebra) \cite{feigin2011quantum,Feigin_2012,Ding:1996mq,Miki2007}.
In this section, we give the definition of the algebra, show the existence of the Hopf-algebra structure, and construct one special vertical representation\footnote{For a comprehensive review, see \cite{Awata_2019} for instance.}. All of these properties have analogs in the quiver quantum toroidal algebra and will be examined later. The readers who are familiar with these materials can skip this section.

\subsection{Definition}
The algebra is defined by two independent parameters. To make it symmetric, we denote the parameters as $q_{1},q_{2},q_{3}$, under the condition $q_{1}q_{2}q_{3}=1$.
The generators of the algebra are
\begin{align}
    E(z) = \sum_{k\in \mathbb{Z}} E_k z^{-k},\quad F(z) = \sum_{k\in \mathbb{Z}} F_k z^{-k},\quad K^\pm (z)=K^\pm \exp\left(\pm \sum_{r=1}^\infty H_{\pm r} z^{\mp r}\right),
\end{align}
and a central element $C$.
$K^-=(K^+)^{-1}$ is also a central element.
The analog of the bond factor $\varphi^{1\Rightarrow 1}(u)$ is
\begin{equation}
    \varphi(z,w) = \prod_{i=1}^3 \frac{(q_i^{1/2}z -q_i^{-1/2}w)}{(q_i^{-1/2}z -q_i^{1/2}w)},\label{eq:gzw}
\end{equation}
where we omit the upper index for simplicity.
The degenerate limit to obtain the affine Yangian is to rewrite $q_i=e^{\epsilon h_{11}^{i}}$ and take $\epsilon\to 0$. 

Equation (\ref{eq:gzw}) reduces to the bond factor for the quiver set 
\ref{fig:quiver_gl}(a) if we write $z=e^{\epsilon x}$ and $w=e^{\epsilon y}$.
The loop condition $h^{(1)}_{11}+h^{(2)}_{11}+h^{(3)}_{11}=0$ corresponds to the multiplicative condition 
$q_{1}q_{2}q_{3}=1$.

The defining relations are as follows:
\begin{align}\label{eq:gl1_relations}
\begin{split}
    &K^\pm(z) K^\pm(w) = K^\pm(w) K^\pm(z),\\
    & K^-(z) K^+(w) = \frac{\varphi(z, Cw)}{\varphi(Cz,w)} K^+(w) K^-(z),\\
    & K^\pm (C^{\frac{1\mp 1}{2}}z) E(w)= \varphi(z,w) E(w) K^\pm (C^{\frac{1\mp 1}{2}}z),\\
    & K^\pm (C^{\frac{1\pm 1}{2}} z) F(w)=\varphi(z,w)^{-1} F(w) K^\pm (C^{\frac{1\pm 1}{2}} z),\\
    & [E(z), F(w)] = \delta\left(C\frac{w}{z}\right)K^+(z)-\delta\left(C\frac{z}{w}\right)K^-(w),\\
    & E(z) E(w) = \varphi(z,w) E(w) E(z),\\
    & F(z) F(w) =\varphi(z,w)^{-1} F(w) F(z),\\
    & [E_0, [E_1, E_{-1}]]=0,\\
    & [F_0, [F_1, F_{-1}]]=0,
    \end{split}
\end{align}
where $\delta(z)$ is the delta function
\begin{equation}
    \delta(z) = \sum_{k\in \mathbb{Z}} z^k.
\end{equation}
The last two equations in (\ref{eq:gl1_relations}) are referred to as Serre relations. Generally, determining these relations for other types of quantum toroidal algebras is difficult, and we will not discuss it from now on. 
Except for the modification of the bond factor and the mode expansion of Drinfeld currents, the difference from the affine Yangian of $\mathfrak{gl}_{1}$ is the existence of the central charge $C$.

Both the generators and the relations depend on $q_1, q_2, q_3$ symmetrically, so this algebra has a triality automorphism of exchanging $q_1, q_2, q_3$. There is $SL(2,\mathbb{Z})$ duality automorphism, which was found by Miki \cite{Miki2007}.  While the former is kept in the degenerate limit, the latter does not exist in the affine Yangian. In this sense, the $q$-deformation makes the algebra more symmetric.

\subsection{Hopf algebra structure}\label{sec:gl1_Hopfstruc}
It is known that quantum toroidal $\mathfrak{gl}_{1}$ has a Hopf algebra structure, which is not so clear in the degenerate limit.
A Hopf algebra $H$ is a bialgebra equipped with a unit $1_{H}$, a counit $\epsilon$, a product $m$, a coproduct $\Delta$ and an antipode satisfying the following properties.
\begin{itemize}
    \item $H$ is an associative and coassociative algebra. This implies $m(1\otimes\epsilon )\Delta=m(\epsilon\otimes1)\Delta=1$ and $(\text{id}\otimes \Delta)\Delta=(\Delta\otimes\text{id})\Delta$.
    \item The counit $\epsilon:H\rightarrow \mathbb{C}$ and the coproduct $\Delta:H\rightarrow H\otimes H$ are homomorphisms of the algebra.
    \item The unit $1_{H}:\mathbb{C}\rightarrow H$ and the product $m:H\otimes H\rightarrow H$ are homomorphisms of the algebra.
    \item The antipode $S:H\rightarrow H$ is an antihomomorphism that obeys $m(S\otimes \text{id})\Delta=\epsilon=m(\text{id}\otimes S)\Delta$.
\end{itemize}
Quantum toroidal $\mathfrak{gl}_1$ is equipped with a formal coproduct,
\begin{align}
\begin{split}
    &\Delta E(z)=E(z)\otimes 1+K^{-}(C_{1}z)\otimes E(C_{1}z),\\
&\Delta F(z)=F(C_{2}z)\otimes K^{+}(C_{2}z)+1\otimes F(z),\\
&\Delta K^{+}(z)=K^{+}(z)\otimes K^{+}(C_{1}^{-1}z),\\
&\Delta K^{-}(z)=K^{-}(C_{2}^{-1}z)\otimes K^{-}(z),\\
&\Delta C=C\otimes C,
\end{split}\label{eq:gl1_coproduct}
\end{align}
where $C_{1}=C\otimes1$ and $C_{2}=1\otimes C$.

We can also define the counit and antipode as,
\begin{align}
\begin{split}
&\epsilon(E(z))=\epsilon(F(z))=0,\\
&\epsilon(K^{\pm}(z))=\epsilon(C)=1,\\
&S(E(z))=-(K^{-}(z))^{-1}E(C^{-1}z),\\
&S(F(z))=-F(C^{-1}z)(K^{+}(z))^{-1},\\
&S(K^{\pm}(z))=(K^{\pm}(Cz))^{-1},\\
&S(C)=C^{-1},
\end{split}\label{eq:gl1_Hopf}
\end{align}
where maps $\Delta$ and $\epsilon$ are extended to algebra homomorphisms, and the map $S$ to an algebra anti-homomorphism, $S(xy)=S(y)S(x)$. 

\subsection{Representations}\label{sec:gl1_plane}

The representations of quantum toroidal $\mathfrak{gl}_1$ depend on two central elements, $C$ and $K^-$.
Due to Miki duality \cite{Miki2007}, they are dual to each other. The representation for $C=1$ (resp. $C\neq 1$) is referred to as vertical (resp. horizontal). This section reviews the vertical representation where the plane partition labels the basis. The affine Yangian shares this type of representation since $C$ is irrelevant. The plane partition realization gives the simplest example of the crystal melting representation in \cite{Li:2020rij}. In the following, the central element $K^-$ is identified with the central charge $K^{1/2}$ for simplicity.

\begin{figure}[h]
    \centering
    \includegraphics[width=10cm]{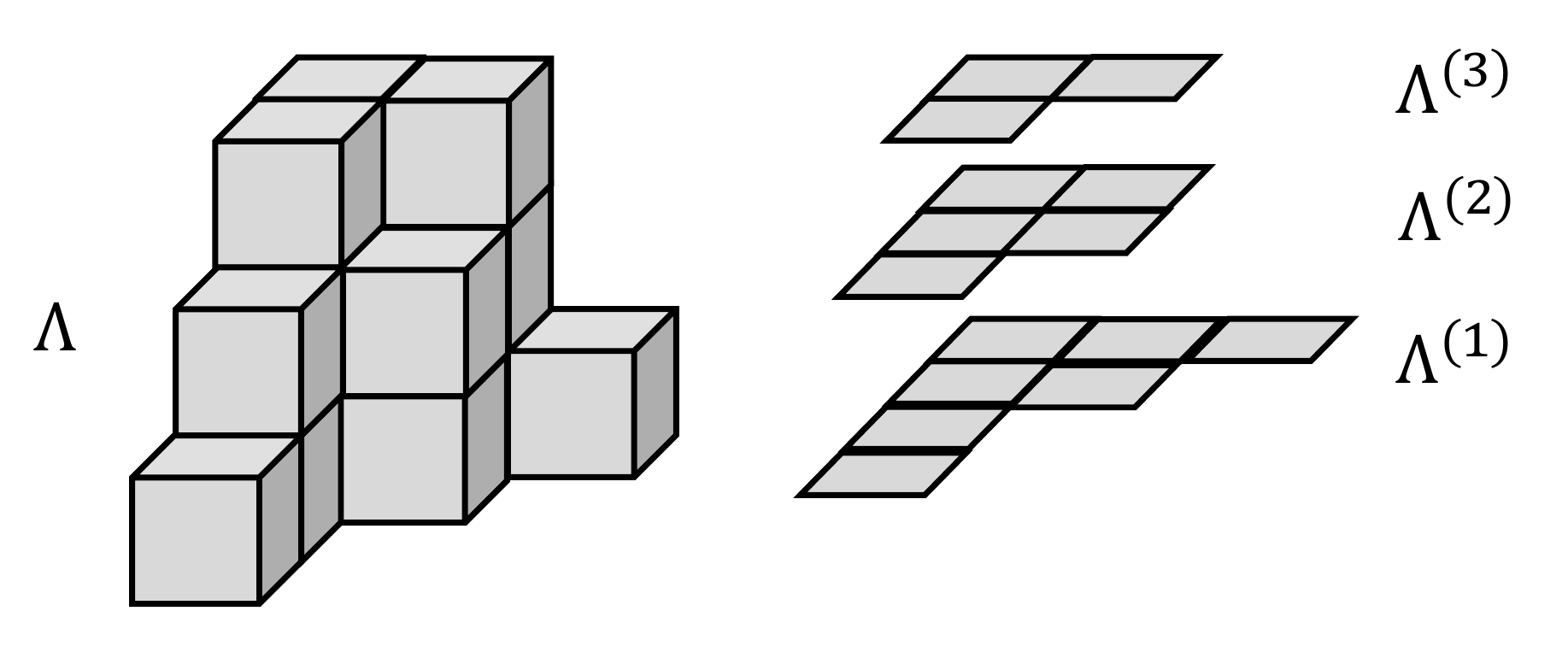}
    \caption{3d Young diagram as a combination of 2d Young diagrams}
    \label{fig:planepartition}
\end{figure}
Plane partition is a set of boxes stacked in a three-dimensional way. Plane partition $\Lambda$ is equivalent to a combination of several Young diagrams $\Lambda^{(k)}$ as in Figure \ref{fig:planepartition},
\begin{equation}
    \Lambda=(\Lambda^{(1)}, \Lambda^{(2)}, \Lambda^{(3)}, \cdots).
\end{equation}
If $j<k$, $\Lambda^{(k)}$ is smaller than or equal to $\Lambda^{(j)}$.

We represent the state labeled by plane partitions as
\begin{equation}
    \mathcal{M}(u, K) = \bigoplus_\Lambda \mathbb{C} \ket{\Lambda},
\end{equation}
where $\Lambda^{(N)}=\emptyset$ if $N$ is large enough.
This state depends on two arbitrary parameters $u$ and $K$.

\begin{figure}[t]
    \begin{tabular}{ccc}
      \begin{minipage}{0.3\hsize}
        \centering
        \includegraphics[width=4cm]{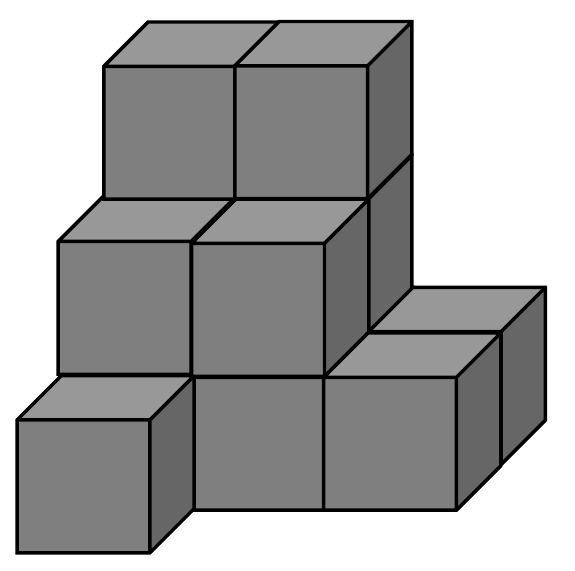}
        \subcaption{$Y_\Lambda$}
      \end{minipage} &
      \begin{minipage}{0.3\hsize}
        \centering
        \includegraphics[width=4cm]{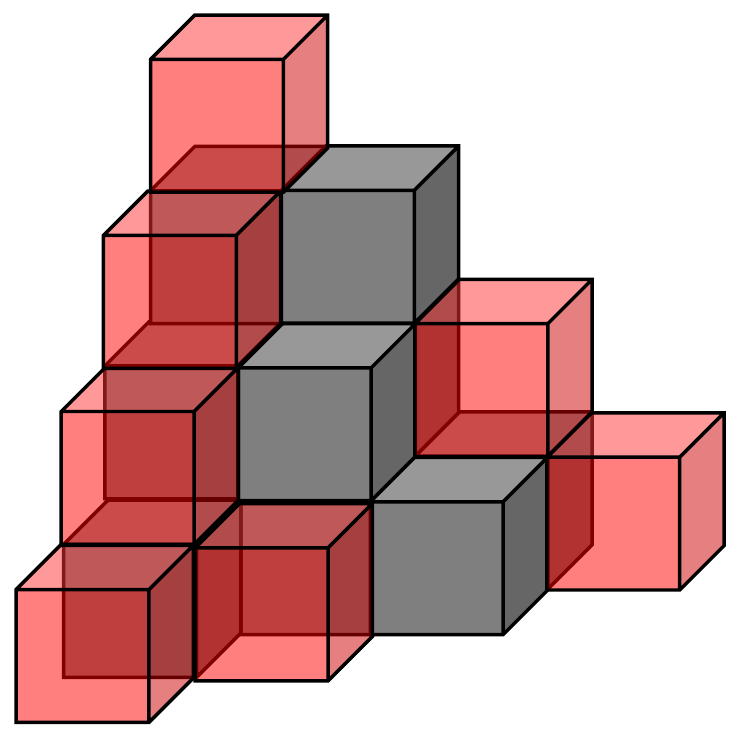}
        \subcaption{$CC(Y_\Lambda)$}
      \end{minipage} &
      \begin{minipage}{0.3\hsize}
        \centering
        \includegraphics[width=4cm]{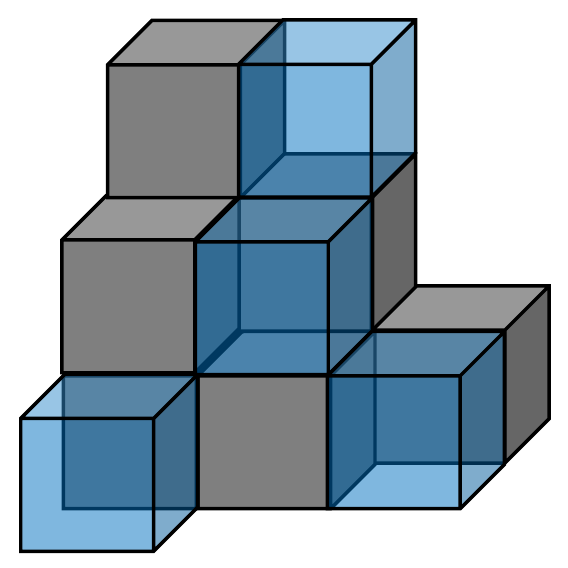}
        \subcaption{$CV(Y_\Lambda)$}
      \end{minipage}
    \end{tabular}
\caption{A configuration of the plane partition. (a) is the plane partition. (b) is the set of places where we can add boxes and (c) is the set of places where we can remove the boxes.}
\label{fig:box_PP}
\end{figure}

The plane partition is characterized as one kind of the 3d crystal determined by the set of a quiver diagram and loop constraints $Q=(Q_0, Q_1, Q_2)$ in Section \ref{sec:QYreview}.
In the case of the quantum toroidal $\mathfrak{gl}_1$, $Q$ is given by (\ref{eq:Q012_gl1}).
There are three parameters $q_1, q_2, q_3$ corresponding to the three arrows, and the $q$-deformation of the constraint \ref{eq:QYloopconstr} is $q_1q_2q_3=1$.

Each path (a set of arrows) on the quiver diagram expresses one box on the plane partition.
The three types of arrows correspond to going to the $x$-, $y$-, and $z$-axis directions in the plane partition, respectively. The loop through each of the three types of arrows once means going from the coordinates $(x,y,z)$ to $(x+1,y+1,z+1)$ on the plane partition, which is one step into the interior from the surface of the plane partition.

We give the actions of the algebra to the basis $\ket{\Lambda}$.
$\ket{\Lambda}$ is the eigenstate of $K^+(z)$ and $K^-(z)$:
\begin{align}
    K^\pm (z)\ket{\Lambda} &= \psi_\Lambda(u/z) \ket{\Lambda},\\
    \psi_\Lambda(u/z) &=\frac{K^{-1/2}-K^{1/2}u/z}{1-u/z}\notag\\
    \times &\prod_{(i,j,k)\in Y_\Lambda}\frac{(1-q_1^j q_2^{k-1} q_3^i u/z) (1-q_1^{j-1} q_2^{k} q_3^i u/z) (1-q_1^j q_2^{k} q_3^{i-1} u/z)}{(1-q_1^{j-1} q_2^{k} q_3^{i-1} u/z) (1-q_1^{j-1} q_2^{k-1} q_3^i u/z) (1-q_1^j q_2^{k-1} q_3^{i-1} u/z)},\label{eq:gl1planeK}
\end{align}
where $Y_\Lambda$ is the set of boxes in the plane partition $\Lambda$. We note that the eigenvalue $\psi_{\Lambda}(u/z)$ should be understood as a formal expansion in $z^{\mp1}$ when $K^{\pm}(z)$.

$E(z)$ adds a box to the basis $\ket{\Lambda}$, and the configuration with an additional box also satisfies the condition of the plane partition (see Figure \ref{fig:box_PP}):
\begin{align}
    E(z)\ket{\Lambda}&=\frac{1}{1-q_1}\sum_{(i,j,k)\in CC(Y_\Lambda)}
    \psi_{\Lambda, i, j, k} \psi_{\Lambda^{(k)}, i} 
    \delta(q_1^j q_2^k q_3^i u/z) \ket{\Lambda+1_i^{(k)}},\label{eq:gl1planeE}\\
    \psi_{\Lambda, i, j, k} &= \prod_{m=1}^{k-1}\psi_{\Lambda^{(m)}}(q_1^{-j}q_2^{m-k-1}q_3^{-i}),\\
    \psi_\lambda(z) &= \prod_{l=1}^\infty 
    \frac{(1-q_1^{\lambda_l-l+1}q_2^{-l+1}q_3/z)(1-q_1^{\lambda_l-l+1}q_2^{-l+2}/z)}{(1-q_1^{\lambda_l-l+1}q_2^{-l+1}/z)(1-q_1^{\lambda_l-l+1}q_2^{-l+2}q_3/z)},\\
    \psi_{\lambda, i}&= \prod_{l=1}^{i-1}\frac{(1-q_1^{\lambda_l-\lambda_i} q_3^{l-i+1})(1-q_1^{\lambda_l-\lambda_i-1} q_3^{l-i-1})}{(1-q_1^{\lambda_l-\lambda_i} q_3^{l-i})(1-q_1^{\lambda_l-\lambda_i-1} q_3^{l-i})},
\end{align}
where $CC(Y_\Lambda)$ is the set of places we can add a box keeping the condition of the plane partition, $\ket{\Lambda+1_i^{(k)}}$ means the basis corresponding to the plane partition with an additional box, and $\lambda$ is any Young diagram.
$F(z)$ removes a box from the basis $\ket{\Lambda}$, and the configuration with the box removed also satisfies the condition of the plane partition:
\begin{align}
    F(z)\ket{\Lambda}&=\frac{q_1}{1-q_1}\sum_{(i,j,k)\in CV(Y_\Lambda)}
    \psi'_{\Lambda, i, j, k} \psi'_{\Lambda^{(k)}, i} 
    \delta(q_1^j q_2^k q_3^i u/z) \ket{\Lambda-1_i^{(k)}},\label{eq:gl1planeF}\\
    \psi'_{\Lambda, i, j, k}&=\lim_{N\to \infty}\prod_{m=k+1}^N \psi_{\Lambda^{(m)}}(q_1^{-j}q_2^{m-k-2}q_3^{-i})\times \frac{K^{-1/2}-K^{1/2}q_1^{-j} q_2^{-k-1} q_3^{-i}}{1-q_1^{-j} q_2^{N-k-1} q_3^{-i}},\\
    \psi'_{\lambda, i}&=\frac{1-q_1^{\lambda_{i+1}-\lambda_i}}{1-q_1^{\lambda_{i+1}-\lambda_i+1}q_3}
    \prod_{l=i+1}^\infty \frac{(1-q_1^{\lambda_l-\lambda_i+1}q_3^{l-i+1})(1-q_1^{\lambda_{l+1}-\lambda_i}q_3^{l-i})}{(1-q_1^{\lambda_{l+1}-\lambda_i+1}q_3^{l-i+1})(1-q_1^{\lambda_l-\lambda_i}q_3^{l-i})},
\end{align}
where $CV(Y_\Lambda)$ is the set of places we can remove a box keeping the condition of the plane partition.

\section{Quiver Quantum Toroidal Algebra}\label{sec:Algebra}
In this section, we define the quiver quantum toroidal algebra (QQTA) $\ddot{\mathcal{U}}_{(Q,W)}$ and summarize its general properties. 

In section \ref{sec:defining_relations}, we give the definition of the algebra and show that it is a natural generalization of the quiver Yangian. In section \ref{sec:associativity} and \ref{sec:Hopf_structure}, we give consistency checks. We show the algebra is an associative algebra and that it has a Hopf superalgebra structure. This Hopf superalgebra structure, especially the coproduct structure, is a property that is not obvious in the quiver Yangian case and will be essential in deriving representations.

\subsection{Conventions}
We define QQTA as a $q$-deformation of the quiver Yangian as the quantum toroidal $\mathfrak{gl}_{1}$ reviewed in the previous section.
It shares the same quiver data as defined in section \ref{sec:quiver-data}
but the bond factor is modified.

We use the same notation of the quiver set in section \ref{sec:quiver-data},
$Q=(Q_{0},Q_{1},Q_{2})$, where $Q_{0}$ is a set of vertices, $Q_{1}$ is a set of arrows between two vertices, and $Q_2$ is a set of closed loops by combining the arrows in $Q_1$. The following notation is used:
\begin{align}
    Q_{0}=\{i\}_{i\in Q_{0}},\quad Q_{1}=\{I\}_{I\in Q_{1}},\quad Q_2=\{L\}_{L\in Q_2},
\end{align}
namely $i,j,..$ are used to label vertices and $I,J,..$ are used to label arrows of the quiver diagram. 

The starting vertex and the ending vertex of the arrow $I$ are denoted by $s(I)$ and $t(I)$, respectively.
A parameter $q_{I}\in \mathbb{C}$ is associated with each arrow $I$ in $Q_1$, which replaces the parameters $h_{I}$ for the quiver Yangian.

For the quiver quantum toroidal algebra, (\ref{eq:QYloopconstr}) is replaced by:
\begin{screen}
\begin{align}
   \text{loop constraint:}\quad \prod_{I\in L}q_{I}=1,\label{eq:loop_cond}
\end{align}
\end{screen}
for each $L\in Q_2$. 
 
As in the quiver Yangian, not all of the loops in $Q_{2}$ are independent, and the number of the independent conditions will be $|Q_{2}|-1$.  

The analog of the vertex constraint will be discussed later (\ref{eq:vertexconstraint}).

\subsection{Generators and Defining Relations}\label{sec:defining_relations}
The QQTA $\ddot{\mathcal{U}}_{(Q,W)}$ is generated by $E_{i,k},F_{i,k},H_{i,r}$, and invertible elements $K_{i},C$, where $i\in Q_{0},k\in\mathbb{Z},r\in\mathbb{Z}^{\times}$.
 The Drinfeld currents are defined by, 
\begin{align}
    E_{i}(z)=\sum_{k\in\mathbb{Z}}E_{i,k}z^{-k},\quad F_{i}(z)=\sum_{k\in\mathbb{Z}}F_{i,k}z^{-k},\quad K_{i}^{\pm}(z)=K_{i}^{\pm1}\exp\left(\pm\sum_{r=1}^{\infty}H_{i,\pm r}z^{\mp r}\right).\label{eq:QQTAgenerator}
\end{align}
The generator associated with the bosonic (resp. fermionic) vertex in $Q_0$ are regarded as the bosonic (resp. fermionic) operator.

The QQTA operator algebra is given as follows:
\begin{screen}
\begin{align}
\begin{split}
    K_{i}K_{i}^{-1}&=K_{i}^{-1}K_{i}=1,\\
    C^{-1}C&=CC^{-1}=1,\\
    K_{i}^{\pm}(z)K_{j}^{\pm}(w)&=K_{j}^{\pm}(w)K_{i}^{\pm}(z),\\
    K_{i}^{-}(z)K_{j}^{+}(w)&=\frac{\varphi^{j\Rightarrow i}(z,Cw)}{\varphi^{j\Rightarrow i}(Cz,w)}K_{j}^{+}(w)K_{i}^{-}(z),\\
    K_{i}^{\pm}(C^{\frac{1\mp1}{2}}z)E_{j}(w)&=\varphi^{j\Rightarrow i}(z,w)E_{j}(w)K_{i}^{\pm}(C^{\frac{1\mp1}{2}}z),\\
    K_{i}^{\pm}(C^{\frac{1\pm1}{2}}z)F_{j}(w)&=\varphi^{j\Rightarrow i}(z,w)^{-1}F_{j}(w)K_{i}^{\pm}(C^{\frac{1\pm1}{2}}z),\\
    [E_{i}(z),F_{j}(w)]=\delta_{i,j}&\left(\delta\left(\frac{Cw}{z}\right)K_{i}^{+}(z)-\delta\left(\frac{Cz}{w}\right)K_{i}^{-}(w)\right),\\
    E_{i}(z)E_{j}(w)&=(-1)^{|i||j|}\varphi^{j\Rightarrow i}(z,w)E_{j}(w)E_{i}(z),\\
    F_{i}(z)F_{j}(w)&=(-1)^{|i||j|}\varphi^{j\Rightarrow i}(z,w)^{-1}F_{j}(w)F_{i}(z),
\end{split}\label{eq:defofQuiverAlgebra}
\end{align}
\end{screen}
The commutator above must be understood in the usual superalgebra sense. When both operators are fermionic, it is an anti-commutator. Otherwise, it is a commutator. 

The bond factor $\varphi^{i\Rightarrow j}(z,w)$ is defined to be 
\begin{align}
    \varphi^{i\Rightarrow j}(z,w)=\frac{\prod_{I\in\{j\rightarrow i\}}(q_{I}^{1/2}z-q_{I}^{-1/2}w)}{\prod_{I\in\{i\rightarrow j\}}(q_{I}^{-1/2}z-q_{I}^{1/2}w)}=\frac{\prod_{I\in\{j\rightarrow i\}}\phi(q_{I};z,w)}{\prod_{I\in\{i\rightarrow j\}}\phi(q_{I}^{-1};z,w)},\label{eq:defstruc}
\end{align}
where 

\begin{align}
    \phi(p;z,w)=p^{1/2}z-p^{-1/2}w.
\end{align} 

When there are no arrows between the two vertices the bond factor is trivial:
\begin{equation}
    \varphi^{i\Rightarrow j}(z,w)=1.
\end{equation}

\paragraph{Properties of bond factors}
The bond factor (\ref{eq:defstruc}) has the following property:
\begin{align}
    \varphi^{i\Rightarrow j}(az,w)=a^{|j\rightarrow i|-|i\rightarrow j|}\varphi^{i\Rightarrow j}(z,a^{-1}w).\label{eq:no4cycle_bondfactor_symmetry}
\end{align}
For the symmetric quiver set, this becomes  $\varphi^{i\Rightarrow j}(az,w)=\varphi^{i\Rightarrow j}(z,a^{-1}w)$.

We note that the above operator algebras should be understood as the relations between the coefficients of the formal power expansion in $z$ and $w$.
For example, the right-hand side of the $K^{+}E$ relation should be understood as the expansion in terms $z^{-1}$ since $K^{+}_{i}(z)$ is expanded in that way. The expansion of the bond factor should be computed by
$\varphi^{i\Rightarrow j}(z,w)=\varphi^{i\Rightarrow j}(1,w/z)$
for the symmetric quiver set,
and the order of $z$ of both sides matches. 
For the asymmetric case, such a simplification does not hold, and we must carefully interpret the defining relations. In this paper, we will focus on the symmetric case.

We expect that the quiver quantum toroidal algebra defined above becomes the quiver Yangian in the degenerate limit. One piece of evidence is obtained by comparing the bond factors. 
The relation between the parameters $\{q_{I}\}$ and the parameters $\{h_{I}\}$ in section \ref{sec:QYreview} or \cite{Li:2020rij} is 
\begin{equation}
    q_{I}=e^{\epsilon h_{I}},\label{eq:param_exp}
\end{equation}
where $\epsilon$ is an infinitesimal parameter. 
The spectral parameters $z$ and $w$ can be written
\begin{equation}
    z=e^{\epsilon x},\quad w=e^{\epsilon y}.
\end{equation}
In the limit $\epsilon\rightarrow 0$, using $q_{I}\sim 1+\epsilon h_{I}$, $z\sim 1+\epsilon x$ and $w\sim 1+\epsilon y$ we obtain 
\begin{align}
\begin{split}
 \varphi^{i\Rightarrow j}(z,w)&=\frac{\prod_{I\in\{j\rightarrow i\}}(q_{I}^{1/2}z-q_{I}^{-1/2}w)}{\prod_{I\in\{i\rightarrow j\}}(q_{I}^{-1/2}z-q_{I}^{1/2}w)}\\
 &\sim \frac{\prod_{I\in\{j\rightarrow i\}}(x-y+h_{I})}{\prod_{I\in\{i\rightarrow j\}}(x-y-h_{I})}, 
 \end{split}
 \end{align}
 where the right-hand side of the last equation is (\ref{eq:QYbondfactor}).

In this sense, the QQTA may be regarded as a $q$-deformation of the quiver Yangian.  

\paragraph{Mode expansions and central element}
The algebra above has an additional central element. Some of the defining relations (\ref{eq:defofQuiverAlgebra}) can be written as follows:
\begin{align}
\begin{split}
 K_{i}K_{j}&=K_{j}K_{i},\\
    [H_{i,r},H_{j,s}]&=\delta_{r+s,0}\frac{C^{r}-C^{-r}}{r}(\sum_{I\in\{j\rightarrow i\}}q_{I}^{r}-\sum_{I\in\{i\rightarrow j\}}q_{I}^{-r}),\\
    K_{i}E_{j}(w)K_{i}^{-1}&=\prod_{\substack{I\in\{i\rightarrow j \}\\ J\in\{j\rightarrow i\}}}q_{I}^{1/2}q_{J}^{1/2}E_{j}(w),\quad \quad K_{i}F_{j}(w)K_{i}^{-1}=\prod_{\substack{I\in\{i\rightarrow j \}\\ J\in\{j\rightarrow i\}}}q_{I}^{-1/2}q_{J}^{-1/2}F_{j}(w),\\
    [H_{i,r},E_{j}(z)]&=\frac{C^{(r-|r|)/2}z^{r}}{r}(\sum_{I\in\{j\rightarrow i\}}q_{I}^{r}-\sum_{I\in\{i\rightarrow j\}}q_{I}^{-r})E_{j}(z),\\
   [H_{i,r},F_{j}(z)]&=-\frac{C^{(r+|r|)/2}z^{r}}{r}(\sum_{I\in\{j\rightarrow i\}}q_{I}^{r}-\sum_{I\in\{i\rightarrow j\}}q_{I}^{-r})F_{j}(z).
\end{split}\label{eq:modeexp}
\end{align}
 We can define a central element as
\begin{align}
    \kappa\equiv\prod_{i\in Q_{0}}K_{i}^{-1}.\label{eq:centralelement}
\end{align}

This element indeed commutes with the other generators. For instance, one obtains the following commutation relations from (\ref{eq:modeexp}):
\begin{align}
\begin{split}
    &\kappa^{-1} E_{j}(z)\kappa=\prod_{i\in Q_{0}}\prod_{\substack{I\in\{i\rightarrow j\}\\J\in\{j\rightarrow i\}}}q_{I}^{1/2}q_{J}^{1/2} E_{j}(z)=\prod_{I\in j}q_{I}^{1/2}E_{j}(z),\\
    &\kappa^{-1} F_{j}(z)\kappa=\prod_{i\in Q_{0}}\prod_{\substack{I\in\{i\rightarrow j\}\\J\in\{j\rightarrow i\}}}q_{I}^{-1/2}q_{J}^{-1/2} E_{j}(z)=\prod_{I\in j}q_{I}^{-1/2}F_{j}(z),
    \end{split}
\end{align}
where $\prod_{I\in j}$ means the product of all arrows going in or out of vertex $j$.
The condition for $\kappa$ to commute with the generators is 
\begin{align}
    \prod_{I\in j}q_{I}=1.
\end{align}
This is automatically satisfied from the loop condition (\ref{eq:loop_cond}) for the symmetric quiver set.

\paragraph{Serre relations}
The above algebra should be equipped with a set of Serre relations. This is the same situation as the quiver Yangian case. Serre relations for quantum toroidal $\mathfrak{gl}_{1},\mathfrak{gl}_{n}$, and $\mathfrak{gl}_{m|n}$ are already known (see for instance, \cite{feigin2011quantum,feigin2013representations,bezerra2019quantum}). There are some discussions in  \cite{Li:2020rij} for the quiver Yangian case, but it remains as an open issue. In this paper, we do not touch the Serre relations.

\subsection{Associativity}\label{sec:associativity}
In this subsection, we check the consistency of the defining relations (\ref{eq:defofQuiverAlgebra}).
One of the consistency conditions of the algebra is the associativity. It becomes nontrivial when both $E_{i}(z)$ and $F_{j}(z)$ are included. We consider the product $E_{i}(x)E_{j}(y)F_{k}(z)$ and use the defining relations (\ref{eq:defofQuiverAlgebra}) in two different orders: 
\begin{align}
\begin{split}
    &E_{i}(x)(E_{j}(y)F_{k}(z))\\
    =&E_{i}(x)\left\{(-1)^{|j||k|}F_{k}(z)E_{j}(y)+\delta_{i,j}\left(\delta\left(\frac{Cz}{y}\right)K_{j}^{+}(y)-\delta\left(\frac{Cy}{z}\right)K_{j}^{-}(z)\right) \right\}\\
    =&(-1)^{|j||k|}\left\{(-1)^{|j||k|}F_{k}(z)E_{i}(x)E_{j}(y)+\delta_{i,k}\left(\delta\left(\frac{Cz}{x}\right)K_{i}^{+}(z)-\delta\left(\frac{Cx}{z}\right)K_{i}^{-}(z)\right)E_{j}(y) \right\}\\
    &+\delta_{jk}\left(\delta\left(\frac{Cz}{y}\right)E_{i}(x)K_{j}^{+}(y)-\delta\left(\frac{Cy}{z}\right)E_{i}(x)K_{j}^{-}(z)\right)\\
    =&(-1)^{|i||j|+|i||k|+|j||k|}\varphi^{j\Rightarrow i}(x,y)F_{k}(z)E_{j}(y)E_{i}(x)\\
    &+(-1)^{|j||k|}\varphi^{j\Rightarrow i}(x,y)\delta_{i,k}\left(\delta\left(\frac{Cz}{x}\right)E_{j}(y)K_{i}^{+}(x)-\delta\left(\frac{Cx}{z}\right)E_{j}(y)K_{i}^{-}(Cx)\right)\\
    &+\delta_{j,k}\varphi^{i\Rightarrow j}(y,x)^{-1}\left( \delta\left(\frac{Cz}{y}\right)K_{j}^{+}(y)E_{i}(x)-\delta\left(\frac{Cy}{z}\right)K_{i}^{-}(Cx)E_{j}(y)\right)
    \end{split}
\end{align}

\begin{align}
\begin{split}
    &(E_{i}(x)E_{j}(y))F_{k}(z)\\
    =&(-1)^{|i||j|}\varphi^{j\Rightarrow i}(x,y)E_{j}(y)(E_{i}(x)F_{k}(z))\\
    =&(-1)^{|i||j|}\varphi^{j\Rightarrow i}(x,y)E_{j}(y)\left\{(-1)^{|j||k|}F_{k}(z)E_{i}(x)+\delta_{i,k}\left(\delta\left(\frac{Cz}{x}\right)K_{i}^{+}(z)-\delta\left(\frac{Cx}{z}\right)K_{i}^{-}(z) \right)  \right\}\\
    =&(-1)^{|i||j|+|i||k|+|j||k|}\varphi^{j\Rightarrow i}(x,y)F_{k}(z)E_{j}(y)E_{i}(x)\\
    &+(-1)^{|j||i|}\varphi^{j\Rightarrow i}(x,y)\delta_{i,k}\left(\delta\left(\frac{Cz}{x}\right)E_{j}(y)K_{i}^{+}(x)-\delta\left(\frac{Cx}{z}\right)E_{j}(y)K_{i}^{-}(Cx)\right)\\
    &+\delta_{j,k}\varphi^{j\Rightarrow i}(x,y)\left( \delta\left(\frac{Cz}{y}\right)K_{j}^{+}(y)E_{i}(x)-\delta\left(\frac{Cy}{z}\right)K_{i}^{-}(Cx)E_{j}(y)\right)
\end{split}
\end{align}
Therefore the condition for associativity is 
\begin{align}
    \varphi^{j\Rightarrow i}(x,y)\varphi^{i\Rightarrow j}(y,x)=1.\label{eq:assoc}
\end{align}
Other cases can be done in the same way and we obtain the same condition.
Inserting the definition of the bond factor (\ref{eq:defstruc}), we obtain
\begin{align}
\frac{\prod_{I\in\{i\rightarrow j\}}(q_{I}^{1/2}x-q_{I}^{-1/2}y)}{\prod_{I\in\{j\rightarrow i\}}(q_{I}^{-1/2}x-q_{I}^{1/2}y)}\frac{\prod_{I\in\{j\rightarrow i\}}(q_{I}^{1/2}y-q_{I}^{-1/2}x)}{\prod_{I\in\{i\rightarrow j\}}(q_{I}^{-1/2}y-q_{I}^{1/2}x)}=(-1)^{|j\rightarrow i|+|i\rightarrow j|}=1.\label{eq:assoc_no4cycle}
\end{align}
This condition holds automatically for the symmetric case. For the asymmetric case, the associativity condition depends whether the factor $|j\rightarrow i|+|i\rightarrow j|$ is even or not. This only affects a subclass of the toric Calabi-Yau three-folds including compact 4-cycles. One way to resolve this problem is to add a minus factor to the bond factor (see section \ref{sec:cpt4cycle}).\footnote{See also the footnote on page 18 of the v3 of \cite{Li:2020rij}.} In this paper, we will focus only on the symmetric quiver case. 

\subsection{Hopf superalgebra structure}\label{sec:Hopf_structure}

The algebra (\ref{eq:defofQuiverAlgebra}) has a Hopf superalgebra structure as in section \ref{sec:gl1_Hopfstruc}, which is a property we do not have for the quiver Yangian case. 
If the algebra is a superalgebra we have to give a $\mathbb{Z}_{2}$ grading to the algebra and define the mappings properly in section \ref{sec:gl1_Hopfstruc}. 
For example, multiplication $m$ for elements of tensor products must be defined as
\begin{align}
    m\left((x\otimes y), (z\otimes w)\right)=(-1)^{|y||z|}xz\otimes yw.
\end{align}
Using this, we can check that the algebra (\ref{eq:defofQuiverAlgebra}) is equipped with a formal coproduct:
\begin{align}
\begin{split}
    &\Delta E_{i}(z)=E_{i}(z)\otimes 1+K_{i}^{-}(C_{1}z)\otimes E_{i}(C_{1}z),\\
&\Delta F_{i}(z)=F_{i}(C_{2}z)\otimes K_{i}^{+}(C_{2}z)+1\otimes F_{i}(z),\\
&\Delta K_{i}^{+}(z)=K_{i}^{+}(z)\otimes K_{i}^{+}(C_{1}^{-1}z),\\
&\Delta K_{i}^{-}(z)=K_{i}^{-}(C_{2}^{-1}z)\otimes K_{i}^{-}(z),\\
&\Delta C=C\otimes C,
\end{split}\label{eq:coproduct}
\end{align}
where $C_{1}=C\otimes1$ and $C_{2}=1\otimes C$.

We can also define the counit and antipode as,
\begin{align}
\begin{split}
&\epsilon(E_{i}(z))=\epsilon(F_{i}(z))=0,\\
&\epsilon(K_{i}^{\pm}(z))=\epsilon(C)=1,\\
&S(E_{i}(z))=-(K_{i}^{-}(z))^{-1}E_{i}(C^{-1}z),\\
&S(F_{i}(z))=-F_{i}(C^{-1}z)(K_{i}^{+}(z))^{-1},\\
&S(K_{i}^{\pm}(z))=(K_{i}^{\pm}(Cz))^{-1},\\
&S(C)=C^{-1},
\end{split}\label{eq:Hopf}
\end{align}
where maps $\Delta$ and $\epsilon$ are extended to algebra homomorphisms, and the map $S$ to a superalgebra anti-homomorphism, $S(xy)=(-1)^{|x||y|}S(y)S(x)$. One can check that  (\ref{eq:coproduct}) and (\ref{eq:Hopf}) are well defined for the symmetric case by direct calculation. 

In the later sections, we will give a bottom-up approach and construct the algebra by looking at the action on three-dimensional BPS crystals, which means they are representations of the algebra with the central charge $C=1$. Although the existence of $C$ and where it enters in the defining relations are still conjectures, since the algebra defined above has a Hopf superalgebra structure, we expect this algebra still has a meaning. Deriving the algebra, including the central charge $C$ from general discussions, will be postponed for future work.

\subsection{Symmetries}\label{sec:symmetry}
We list down some general symmetries of the algebra. Using these symmetries, we will see we can impose additional conditions on the parameters assigned to the arrows of the quiver diagram.
\subsubsection{Rescaling symmetry of spectral parameter }
One can rescale the spectral parameter by an overall constant:
\begin{align}
\begin{split}
    &z\rightarrow z'=az,\\
    &E_{i}(z)\rightarrow E_{i}'(z)=E_{i}(z'),\\
    &F_{i}(z)\rightarrow F_{i}'(z)=F_{i}(z'),\\
    &K_{i}^{\pm}(z)\rightarrow {K'_{i}}^{\pm}(z)=K_{i}^{\pm}(z').
    \end{split}\label{eq:rescalesymm}
\end{align}
This transformation does not change the algebra and gives an automorphism. Under this transformation, the bond factors transform as
\begin{align}
    \varphi^{i\Rightarrow j}(az,aw)=\varphi^{i\Rightarrow j}(z,w),
\end{align}
which holds for the symmetric quiver set.
In terms of mode generators, the symmetry is 
\begin{align}
    \begin{split}
        E'_{i,k}=a^{-k}E_{i,k},\\
        F'_{i,k}=a^{-k}F_{i,k},\\
        H_{i,\pm r}=a^{\mp r}H_{i,\pm r}.
    \end{split}
\end{align}
\subsubsection{Gauge symmetry shift}
We also have an analog of  ``gauge symmetry'' in \cite{Li:2020rij}. For each vertex, we rescale the parameters $q_{I}$ as
\begin{align}
    q_{I}\rightarrow q_{I}'=q_{I}p_{i}^{\text{sign}_{i}(I)}
\end{align}\label{eq:gaugetransf}
where
\begin{displaymath}
    \text{sign}_{i}(I)\equiv
    \begin{dcases}
    +1\quad(s(I)=i,\quad t(I)\neq i),\\
    -1\quad(s(I)\neq i,\quad t(I)= i),\\
    0\quad(\text{otherwise})
    \end{dcases}
\end{displaymath}
and $p_{i}$ are arbitrary parameters associated to each vertex. We can see this is indeed a symmetry using the rescaling symmetry of the spectral parameters. We consider the case only when we use this gauge transformation on one vertex $i$. The $q_{I}$ dependent part is the bond factor $\varphi^{i\Rightarrow j}(z,w)$. We divide the bond factors into nonzero mode part and zero mode part as
\begin{align}
    \varphi^{i\Rightarrow j}(z,w)=\prod_{I\in\{j\rightarrow i\}}q_{I}^{1/2}\prod_{J\in\{i\rightarrow j\}}q_{J}^{1/2}\frac{\prod_{\{j\rightarrow i\}}\left(1-q_{I}^{-1}\frac{w}{z}\right)}{\prod_{\{i\rightarrow j\}}\left(1-q_{I}\frac{w}{z}\right)}.
\end{align}
Under (\ref{eq:gaugetransf}) the zero mode part changes as 
\begin{align}
    \prod_{I\in\{j\rightarrow i\}}{q'_{I}}^{1/2}\prod_{J\in\{i\rightarrow j\}}{q'_{J}}^{1/2}=\prod_{I\in\{j\rightarrow i\}}q_{I}^{1/2}\prod_{J\in\{i\rightarrow j\}}q_{J}^{1/2}.
\end{align}
When $s(I)=t(I)=i$, it is trivial and when $s(I)\neq t(I)$, we get 
\begin{align}
    \prod_{I\in\{j\rightarrow i\}}{q'_{I}}^{1/2}=\prod_{I\in\{j\rightarrow i\}}q_{I}^{1/2}\prod_{I\in \{j\rightarrow i\}}p_{i}^{-1}\label{eq:vertexconstr_cpt4cond_1}\\
    \prod_{J\in\{i\rightarrow j\}}{q'_{J}}^{1/2}=\prod_{J\in\{i\rightarrow j\}}q_{J}^{1/2}\prod_{I\in\{i\rightarrow j\}}p_{i}^{+1}.\label{eq:vertexconstr_cpt4cond_2}
\end{align}
Since we are considering the symmetric case, $|i\rightarrow j|=|j\rightarrow i|$, we obtain 
\begin{align}
    \prod_{I\in \{j\rightarrow i\}}p_{i}^{-1}\prod_{I\in\{i\rightarrow j\}}p_{i}^{+1}=1.\label{eq:vertexconstr_cpt4cond_3}
\end{align}
The nonzero mode part changes as 
\begin{align}
    \frac{\prod_{\{j\rightarrow i\}}\left(1-{q'_{I}}^{-1}\frac{w}{z}\right)}{\prod_{\{i\rightarrow j\}}\left(1-{q'_{I}}\frac{w}{z}\right)}=\frac{\prod_{\{j\rightarrow i\}}\left(1-{q_{I}}^{-1}p_{i}\frac{w}{z}\right)}{\prod_{\{i\rightarrow j\}}\left(1-{q_{I}}p_{i}\frac{w}{z}\right)}.
\end{align}
These bond factors enter in the defining relations including the generators $E_{i}(z)$, $F_{i}(z)$ and $K_{i}^{\pm}(z)$. Using (\ref{eq:gaugetransf})
we can redefine these generators associated to vertex $i$ by rescaling the spectral parameter. 

For example, the defining relation
\begin{align}
    E_{i}(z)E_{j}(w)=(-1)^{|i||j|}\varphi^{j\Rightarrow i}(z,w)E_{j}(w)E_{i}(z)
\end{align}
will become 
\begin{align}
     E_{i}(z)E_{j}(w)=(-1)^{|i||j|}\prod_{I\in\{i\rightarrow j\}}q_{I}^{1/2}\prod_{J\in\{j\rightarrow i\}}q_{J}^{1/2}\frac{\prod_{\{i\rightarrow j\}}\left(1-q_{I}^{-1}p_{i}^{-1}\frac{w}{z}\right)}{\prod_{\{j\rightarrow i\}}\left(1-q_{I}p_{i}^{-1}\frac{w}{z}\right)}E_{j}(w)E_{i}(z)
\end{align}
under transformation
and by setting $E'_{i}(z)=E_{i}(p_{i}^{-1}z)$ we obtain the same EE relation:
\begin{align}
     E'_{i}(z)E_{j}(w)=(-1)^{|i||j|}\varphi^{j\Rightarrow i}(z,w)E_{j}(w)E'_{i}(z).
\end{align}
We can eliminate this gauge symmetry by imposing gauge fixing conditions such as 
\begin{screen}
\begin{align}
    \text{vertex constraint:}\quad \prod_{I\in Q_{1}(i)}q_{I}^{\text{sign}_{i}(I)}=1.\label{eq:vertexconstraint}
\end{align}
\end{screen}
One can obtain the same constraint by using (\ref{eq:param_exp}) and (\ref{eq:QYvertexconstr}), namely, by exponentiating the vertex constraint of the quiver Yangian.

In this paper, we will always impose this condition. By similar discussions as in \cite{Li:2020rij}, we can see that after imposing (\ref{eq:loop_cond}) and (\ref{eq:vertexconstraint}) we only have two independent parameters.

\section{Bootstrapping Quiver Quantum Toroidal Algebras when \texorpdfstring{$C=1$}{C=1}}\label{sec:bootstrap}
In this section, we bootstrap the algebras (\ref{eq:defofQuiverAlgebra}) when $C=1$ following the strategy of \cite{Li:2020rij} by assuming that the algebras act on the three-dimensional BPS crystals of \cite{Ooguri_2009}.\footnote{We give a brief review of the toric diagram and the associated BPS crystal in appendix \ref{sec:3d_crystal} to make the paper self-contained.} These will be representations of (\ref{eq:defofQuiverAlgebra}) after setting $C=1$. One can see from (\ref{eq:modeexp}) that when $C=1$, $H_{i,r}$ commutes with each other, and $K_{i}(z)\; (i\in Q_{0})$ become simultaneously diagonalizable in this representation. We restrict the discussion to the symmetric quiver set.

The basic strategy in \cite{Li:2020rij} is the following four steps:
\begin{enumerate}
  \item Give an ansatz on the action of the generators on the basis labeled by the three-dimensional crystals. The ansatz is based on three pieces of fundamental information: the pole structure, the moduli of coefficients, and signs of the coefficients. The first two will be determined by the charge function, and the last one will be determined after the algebra is fixed.
  \item Determine the charge function from the quiver diagram by using the ansatz above. The moduli of coefficients will be determined in this process.
  \item Fix the algebra from the quiver data and the ansatz above. The statistics of the operators will be determined in this process.
  \item Fix the signs from the statistics of the algebra.
\end{enumerate}
The essential part of this representation is the eigenvalue of the generators $K^{\pm}_{i}(z)$. We derive the charge functions in section \ref{sec:ansatz}, \ref{sec:coordinate_function}, and \ref{sec:charge_function}. The defining relations of EE and EF are discussed in section \ref{sec:EE,EFrelation}. Other defining relations are in appendix \ref{sec:appendix_bootstrapp}. In section \ref{sec:cpt4cycle}, we summarize where we use the condition of the symmetric quiver set to illuminate the future analysis of the asymmetric case and give some comments.

\subsection{3d crystal}
We construct a 3d crystal from a quiver diagram and loop constraints based on \cite{Ooguri_2009}, which generalizes the plane partition for $\mathfrak{gl}_{1}$.

\begin{figure}[t]
	\centering
	\includegraphics[width=4.5cm]{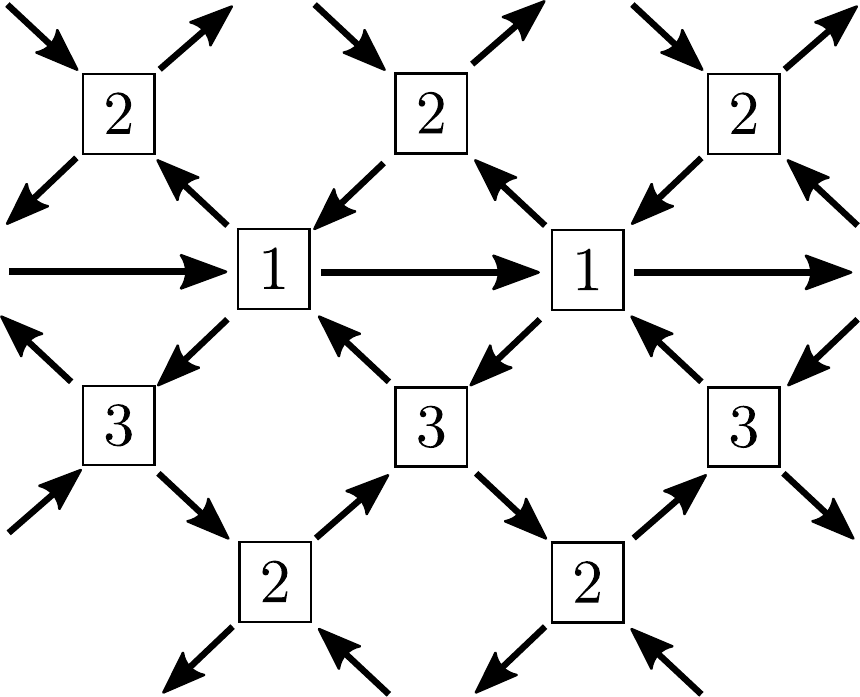}
	\caption{Periodic quiver diagram for $\mathfrak{gl}_{2|1}$. We can determine it from the quiver diagram and $Q_2$ by requiring each loop in $Q_2$ to be a loop on the periodic quiver. In this example, $Q_2$ consists of $1\to 1\to 3\to 1$, $1\to 2\to 3\to 2\to 1$, $1\to 1\to 2\to 1$, $1\to 3\to 2\to 3\to 1$.}
	\label{fig:gl21_periodic}
\end{figure}

First, we extend the quiver diagram to the periodic quiver diagram by using the loop constraints.
By duplicating the vertices, we create a diagram with up to one arrow between each pair of vertices.
From the quiver diagram alone, there are many ways to do such extension, but we can determine uniquely by requiring each loop constraint to be a loop on the periodic quiver.
This is the same as the brane tiling with vertices connected in Appendix \ref{sec:3d_crystal}.
For example, the periodic quiver diagram of $\mathfrak{gl}_{2|1}$ corresponding to Figure \ref{fig:quiver_gl}(b) is shown in Figure \ref{fig:gl21_periodic}.

We choose a vertex on the periodic quiver diagram and set it as the origin. The 3d crystal consists of ``atoms" where each atom corresponds to a family of paths from the origin to a point in 3d crystal.

\begin{figure}[t]
	\centering
	\includegraphics[width=9cm]{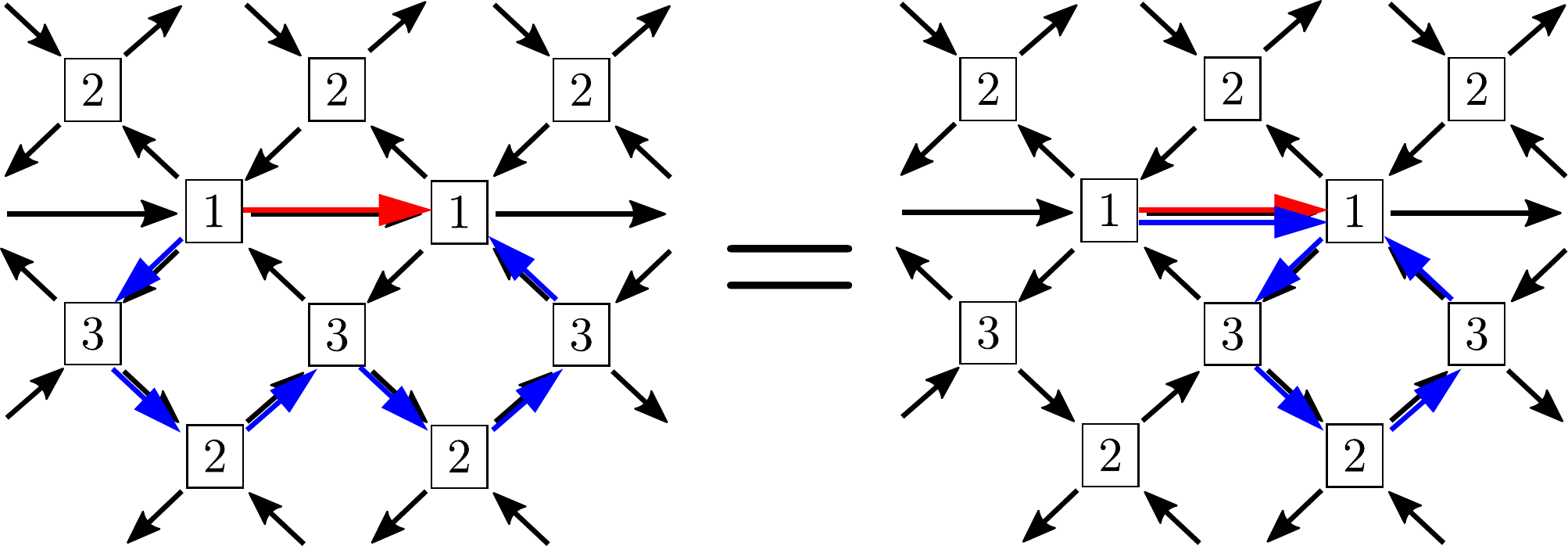}
	\caption{Identification on periodic quiver diagrams. The red path is the shortest path connecting such two points. The two blue paths are identified by the F-term relation, and they are the red shortest path plus one loop.}
	\label{fig:gl21_fterm}
\end{figure}

We use the following relation for the identification of paths.
\begin{itemize}
	\item F-term relations:
	In the periodic quiver diagram, two loops with one common arrow are identified.
\end{itemize}
One can prove that the arbitrary paths from the origin to a point can be identified with the following special path
\begin{equation}
	p=p_0 \omega^n,
\end{equation}
where $p_0$ is the shortest path from the start point of $p$ to the endpoint of $p$, $\omega$ is one of the arbitrary elements in $Q_2$, and $n$ is the number of loops.
For example, the blue paths in Figure \ref{fig:gl21_fterm} are identified by the F-term relation applied to a path $3\to 1$.
The path on the right figure consists of the red shortest path plus one loop.
We note that the shortest path from the origin to any point on the periodic lattice is unique.
It implies that each atom is specified by a point on the periodic lattice and a non-negative integer $n$. We define the set of atoms with $n=0$ as the surface of the 3d crystal.
The non-negative number $n$ measures the distance from the surface, and we call it the depth in the 3d crystal.

\begin{figure}[t]
	\begin{tabular}{cc}
		\begin{minipage}{0.45\hsize}
			\centering
			\includegraphics[width=5cm]{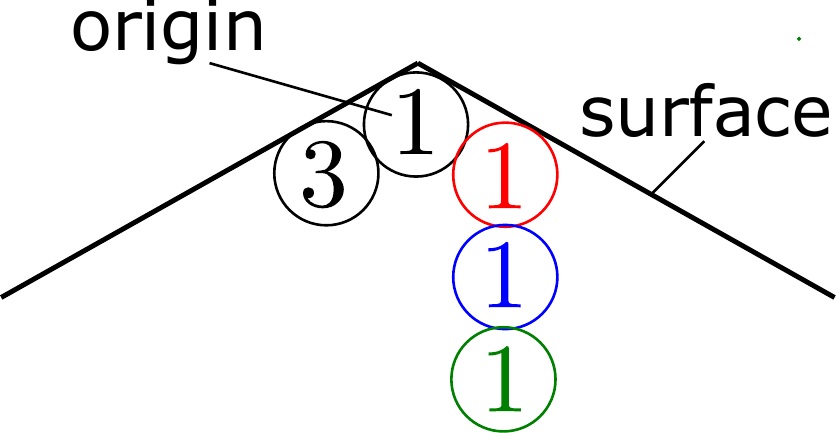}
			\subcaption{Depth of some atoms}
		\end{minipage} &
		\begin{minipage}{0.45\hsize}
			\centering
			\includegraphics[width=4.5cm]{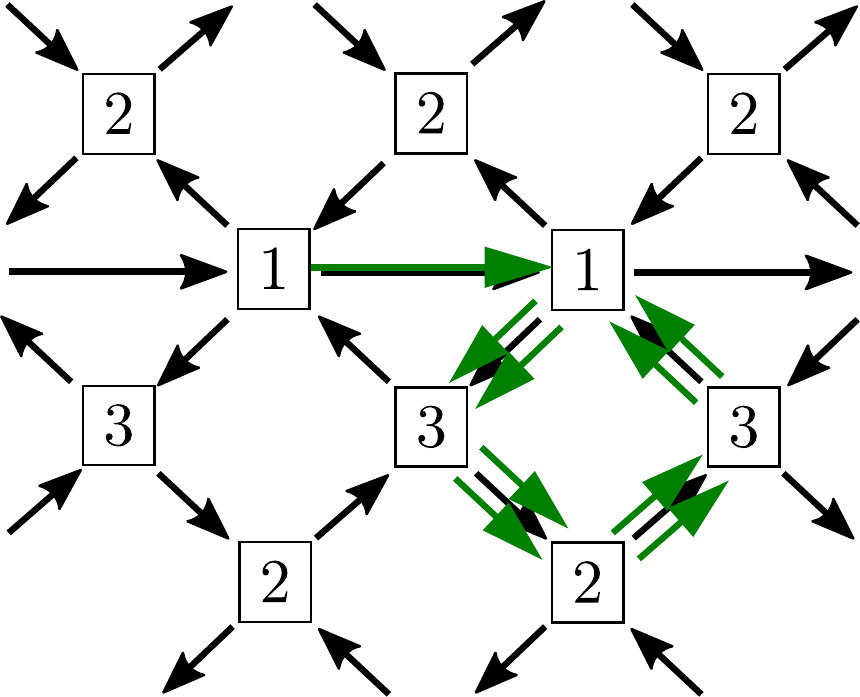}
			\subcaption{A path with three loops}
		\end{minipage}
	\end{tabular}
	\caption{Depth of atoms in a 3d crystal and the number of loops in periodic quiver diagram. The colors in (a) corresponds to those of loops in Figures \ref{fig:gl21_fterm} and \ref{fig:gl21_depth} (b).}
	\label{fig:gl21_depth}
\end{figure}

The red, blue, and green atoms in the left of Figure \ref{fig:gl21_depth} correspond to the red path in Figure \ref{fig:gl21_fterm}, the blue path in Figure \ref{fig:gl21_fterm}, and the green path in the right of Figure \ref{fig:gl21_depth}.
One can find other illuminating figures of the 3d crystal in \cite{Ooguri_2009,Li:2020rij}.

Finally we mention some differences from the plane partition which corresponds to $\mathfrak{gl}_{1}$. In the plane partition case, the periodic quiver diagram consists of a single vertex ``1" and it is connected by three arrows. The surface of 3d crystal is identified with the points $(m_1, m_2, m_3)$ ($m_1, m_2, m_3\geq 0$) with one of $m_i$ vanishes. Two points $(m_1, m_2, m_3)$ and $(m_1+n, m_2+n, m_3+n)$ are connected by $n$ loops $1\xrightarrow{1}1 \xrightarrow{2}1 \xrightarrow{3}1$. By definition, every neighboring boxes (atoms) are connected by arrows. 

\begin{figure}[t]
	\centering
	\includegraphics[width=6cm]{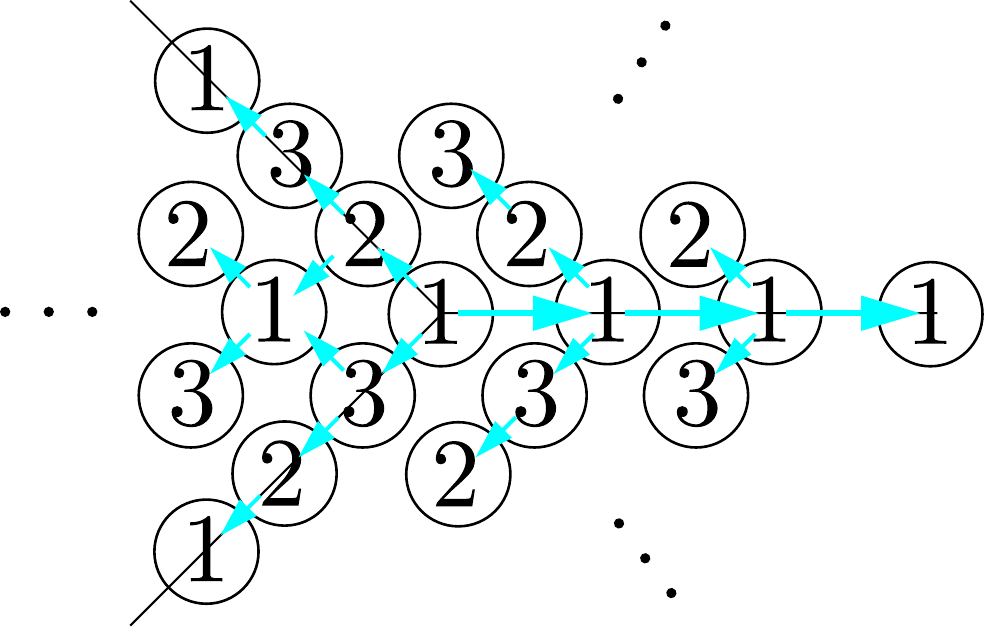}
	\caption{The surface of the 3d crystal for $\mathfrak{gl}_{2|1}$. Light blue arrows express bonds between atoms, and we can see they are different from the plane partition.}
	\label{fig:gl21_surface}
\end{figure}

In the case of the 3d crystal for $\mathfrak{gl}_{2|1}$, each atom has color $1,2,3$. Moreover, the neighboring atoms may not be connected by bonds. See Figure \ref{fig:gl21_surface} which illustrates the atoms on the surface.

\subsection{Ansatz for Representation}\label{sec:ansatz}
We use $\Lambda$ to label a three-dimensional crystal configuration.  The ansatz is a generalization of the quantum toroidal $\mathfrak{gl}_{1}$ plane partition representation in (\ref{eq:gl1planeK}), (\ref{eq:gl1planeE}), and (\ref{eq:gl1planeF}).
\begin{align}
    \begin{split}
        &K_{i}^{\pm}(z)\ket{\Lambda}=[\Psi_{\Lambda}^{(i)}(z,u)]_{\pm}\ket{\Lambda},\\
        &E_{i}(z)\ket{\Lambda}=\sum_{\fbox{$i$}\in\text{Add}(\Lambda)}E^{(i)}(\Lambda\rightarrow \Lambda+\fbox{$i$})\delta\left(\frac{z}{uq(\fbox{$i$})}\right)\ket{\Lambda+\fbox{$i$}},\\
        &F_{i}(z)\ket{\Lambda}=\sum_{\fbox{$i$}\in\text{Rem}(\Lambda)}F^{(i)}(\Lambda\rightarrow \Lambda-\fbox{$i$})\delta\left(\frac{z}{uq(\fbox{i})}\right)\ket{\Lambda-\fbox{$i$}},\label{eq:ansatz}
    \end{split}
\end{align}for $i=1,...|Q_{0}|$, where 
\begin{align}
\begin{split}
    &E^{(i)}(\Lambda\rightarrow \Lambda+\fbox{$i$})=\epsilon(\Lambda\rightarrow \Lambda+\fbox{$i$})\sqrt{p^{(i)}\underset{x=uq(\fbox{$i$})}{\Res}\Psi_{\Lambda}^{(i)}(x,u)},\\
     &F^{(i)}(\Lambda\rightarrow \Lambda-\fbox{$i$})=\epsilon(\Lambda\rightarrow \Lambda-\fbox{$i$})\sqrt{q^{(i)}\underset{y=uq(\fbox{$i$})}{\Res}\Psi_{\Lambda}^{(i)}(y,u)},\\
     &\epsilon(\Lambda\rightarrow \Lambda+\fbox{$i$})=\pm 1,\quad\epsilon(\Lambda\rightarrow \Lambda-\fbox{$i$})=\pm 1. 
     \end{split}\label{eq:ansatzcoefficient}
\end{align}
We note here $[f(z)]_{\pm}$ means formal expansion of $f(z)$ in $z^{\mp}$. We also emphasis that the residue (denoted as $\Res$) here slightly differs from the normal definition. For a rational function $f(z)$ , the residue\footnote{For example, when $f(z)=\frac{z-b}{z-a}$,  $[f(z)]_{+}=\frac{1-b/z}{1-a/z}=\sum_{n=0}^{\infty}(1-\frac{b}{z})(\frac{a}{z})^{n}$ and $[f(z)]_{-}=\frac{z-b}{z-a}=\frac{b}{a}(1-\frac{b}{z})\sum_{n=0}^{\infty}{(\frac{z}{a})^{n}}$, so $[f(z)]_{+}-[f(z)]_{-}=(1-\frac{b}{a})\delta\left(\frac{z}{a}\right)$ and $\underset{z=a}{\Res}f(z)=1-\frac{b}{a}$. The relation between this residue and the normal residue is $\frac{1}{z}{\underset{z=a}{\Res}}^{\text{normal}}f(z)\delta\left(\frac{z}{a}\right)=\underset{z=a}{\Res}^{\text{here}}f(z)\delta\left(\frac{z}{a}\right)$.} is defined as the coefficient of the delta function,
\begin{align}
\sum_{a\in\text{Poles of $f(z)$}}\underset{z=a}{\Res} f(z)\delta(\frac{z}{a})\equiv[f(z)]_{+}-[f(z)]_{-}.\label{eq:residue}
\end{align}

The ansatz (\ref{eq:ansatz}) and (\ref{eq:ansatzcoefficient}) can be understood in the following way.
$K_{i}^{\pm}(z)$ acts diagonally on the three-dimensional crystal configurations and $\Psi_{\Lambda}^{(i)}(z,u)$ is the eigenvalue, which means this representation is a vertical representation. On the other hand, $E_{i}(z)$ adds an atom of color $i$ where it can be added to the crystal, and $F_{i}(z)$ removes an atom of color $i$ where it can be removed from the crystal. Add($\Lambda$) is the set of atoms which can be added to the configuration $\Lambda$, while Rem($\Lambda$) is the set of atoms which can be removed from the configuration $\Lambda$. The charge functions $\Psi_{\Lambda}^{(i)}(z,u)$ determine the pole structure of the algebra. The poles of $\Psi_{\Lambda}^{(i)}(z,u)$ are either the position where the atoms can be added or be removed. The coefficients $E^{(i)}(\Lambda\rightarrow\Lambda+\fbox{$i$})$ and $F^{(i)}(\Lambda\rightarrow\Lambda-\fbox{$i$})$ should show this structure and that is the reason why they are proportional to the residue of the poles. It will be shown later that $p^{(i)}q^{(i)}=\pm1$ and hence we can put $q^{(i)}=1$. We will eventually see that $p^{(i)}$ is related to the statistics of the operators $E^{(i)}(z)$ and $F^{(i)}(z)$.
\subsection{Coordinate function}\label{sec:coordinate_function}
Fix the origin $\mathfrak{o}$ and define the coordinate of \fbox{$i$} as the path from the origin. In the quantum toroidal $\mathfrak{gl}_{1}$ case, this is simply the three-dimensional coordinates of the atom. For general crystals, there are no three-dimensional coordinates, but we can define the coordinate to be the path from the origin in the periodic quiver. In the quiver Yangian case, this was defined as 
\begin{align}
    h(\fbox{$i$})\equiv \sum_{I\in\text{path}[\mathfrak{o}\rightarrow \fbox{$i$}]}h_{I},
\end{align}
namely, the coordinate function for \fbox{$i$} is the sum of all charges along the path from the origin.
For the quantum toroidal version, the coordinates can be defined by changing the sum to product as
\begin{align}
    q(\fbox{$i$})\equiv\prod_{I\in\text{path}[\mathfrak{o}\rightarrow \fbox{$i$}]}q_{I}.
\end{align}
This can be understood as exponentiating the coordinates of the quiver Yangian case (see (\ref{eq:param_exp})).
We note that the loop condition (\ref{eq:loop_cond}) implies that the coordinate function does not depend on the different choice of paths and assigns a unique value to each atom.

In the quantum toroidal case, we can introduce a spectral parameter $u$ to the representation. From now on, the coordinates will be set as \begin{align}z=uq(\fbox{$i$}).\end{align} Using the rescaling symmetry (\ref{eq:rescalesymm}), we can eliminate this and set $u=1$, but we leave it as a generic value.

\subsection{Charge function}\label{sec:charge_function}
Following the degenerate case, we can give the following ansatz for the charge function $\Psi_{\Lambda}^{(i)}(z,u)$, 
\begin{align}
    \Psi_{\Lambda}^{(i)}(z,u)=\psi_{\emptyset}^{(i)}(z,u)\prod_{j\in Q_{0}}\prod_{\fbox{$j$}\in\Lambda}\varphi^{j\Rightarrow i}(z,uq(\fbox{$j$})),\label{eq:chargefunction}
\end{align}
where $\psi_{\emptyset}^{(i)}(z,u)$ is the vacuum charge. Each atom with color $j$ in the configuration $\Lambda$ gives a contribution factor  $\varphi^{j\Rightarrow i}(z,uq(\fbox{$j$}))$ to the charge function of color $i$. To determine the charge function, the basic principle is that the poles of the charge function is either the position where we can add an atom or the position where we can remove an atom. We note that the atoms should satisfy the melting rule of the crystal, and this is ensured under the loop condition (\ref{eq:loop_cond}) (see \cite{Li:2020rij} and \cite{Ooguri_2009} for the melting rule). The charge function can be obtained by considering the action of the algebra on low levels\footnote{Level N means we have N atoms in the crystal configuration.} as in \cite{Li:2020rij}, but we reproduce the discussion here to make it self-contained. 
\subsubsection{Vacuum to Level 1} 
The first atom is in the origin $\mathfrak{o}$ and the coordinate is $q(\fbox{$i$})=1$. We assume the color of the atom in the origin to be $1$. Since the poles of the charge function is either the place of addable atoms or the place of removable atoms, the pole of the vacuum charge is $z=u$ (note that we introduced a spectral parameter $u$). Then we obtain 
\begin{align}
\begin{split}
    \psi_{\emptyset}^{(i)}(z,u)&=\frac{{K^{(i)}}^{-1/2}z-{K^{(i)}}^{1/2}u}{z-u},\\
  K^{(i)}&=(K)^{\delta_{i,1}}.
   \end{split}
\end{align}
When $i\neq 1$ the vacuum charge is  $\psi_{\emptyset}^{(i)}(z,u)=1$ and there is no pole, which means we cannot add this atom to the empty configuration. We can also rewrite this as 
\begin{align}
    \psi_{\emptyset}^{(i)}(z,u)=(\psi_{\emptyset}(z,u))^{\delta_{i,1}},
\end{align}
where \begin{align}
    \psi_{\emptyset}(z,u)=\frac{K^{-1/2}z-K^{1/2}u}{z-u}.
\end{align}
The action of $E^{(i)}$ and $F^{(i)}(z)$ can be written as 
\begin{align}
    \begin{split}
        E_{i}(z)\ket{\emptyset}&=\pm\sqrt{p^{(i)}(1-K^{(i)})}\delta\left(\frac{z}{u}\right)\ket{\fbox{$i$}},\\
        F_{i}(z)\ket{\emptyset}&=0,\\
        K_{i}^{\pm}(z)\ket{\emptyset}&=\left[\frac{{K^{(i)}}^{-1/2}z-{K^{(i)}}^{1/2}u}{z-u}\right]_{\pm}\ket{\emptyset}
    \end{split}\label{eq:level0}
\end{align}
\subsubsection{Level 1 to Level 2}

After setting the atom $1$ at the origin, the configuration now contains only one atom, and the charge function (\ref{eq:chargefunction}) is 
\begin{align}
    \Psi_{\fbox{$1$}}^{(i)}(z,u)=(\psi_{\emptyset}(z,u))^{\delta_{i,1}}\varphi^{1\Rightarrow i}(z,u).
\end{align}
After setting the atom of color $1$ at the origin, the atom we can put in the next level is the atom with colors connected to the vertex $1$ in the quiver diagram. Since the charge function must contain poles at the position where it is possible to add the atom connected to \fbox{$1$}, we obtain
\begin{align}
    \varphi^{1\Rightarrow i}(z,u)\supset\begin{cases}\frac{1}{\prod_{I\in \{1\rightarrow i\}}(q_{I}^{-1/2}z-q_{I}^{1/2}u)}\quad (1\rightarrow i),\\1\quad (1\not\rightarrow i), \end{cases}
\end{align}
where $\{1\rightarrow i\}$ is the set of arrows from vertex $1$ to vertex $i$ in the quiver diagram. 
When $i$ is connected to $1$, then we need to have poles in $z=uq_{I}$ for each arrow $I\in \{1\rightarrow i\}$, which means that all of these poles must be contained in the charge function. On the other hand, for vertices not connected to vertex $1$, we cannot add the atom to the crystal configuration, which means we cannot have any poles. Although we assumed the crystal in the origin to be $1$, we can also do the same thing with other colors. Then we can get 
\begin{align}
    \varphi^{j\Rightarrow i}(z,u)=\frac{N^{j\Rightarrow i}(z,u)}{D^{j\Rightarrow i}(z,u)},\label{eq:chargeansatz}
\end{align}
where 
\begin{align}
    D^{j\Rightarrow i}(z,u)=\prod_{I\in\{j\rightarrow i\}}(q_{I}^{-1/2}z-q_{I}^{1/2}u).\label{eq:chargedenominator}
\end{align}
The action of $K_{i}^{\pm}(z)$ can be written as 
\begin{align}
     K_{i}^{\pm}(z)\ket{\fbox{$1$}}&=\left[(\psi_{\emptyset}(z,u))^{\delta_{i,1}}\varphi^{1\Rightarrow i}(z,u)\right]_{\pm}\ket{\fbox{$1$}}.
\end{align}
The charge function indeed have poles where atoms are addable and removable. We get removing poles at $z=u$ from the vacuum charge function, and we get adding poles from the bond factors. 
The action of $E_{i}(z)$ and $F_{i}(z)$ can be written as 
\begin{align}
    \begin{split}
        E_{i}(z)\ket{\fbox{$1$}}&=\sum_{i\in [1\rightarrow]}\sum_{j}\#\delta\left(\frac{z}{uq(\fbox{$i$}_{j})}\right)\ket{\fbox{$1$}+\fbox{$i$}_{j}},\\
        F_{i}(z)\ket{\fbox{$1$}}&=\delta_{i,1}\#\delta\left(\frac{z}{u}\right)\ket{\emptyset},\label{eq:actionlevel1-2}
    \end{split}
\end{align}
where $[1\rightarrow ]$ is the set of vertices connected to vertex $1$ in the quiver diagram, and the summation with $j$ is took among the atoms of color $i$ addable to the crystal configuration. $E_{i}(z)$ adds atoms whose position are in the poles, while $F_{i}(z)$ removes the atom at the origin. 
\subsubsection{Level 2}
We assume that $\fbox{1}$ is placed at the origin and that $\fbox{$j$}$ is the next atom placed after $\fbox{$1$}$. The charge function of this configuration is 
\begin{align}
\begin{split}
    \Psi_{\Lambda}^{(i)}(z,u)&=\psi_{\emptyset}^{(i)}(z,u)\varphi^{1\Rightarrow i}(z,u)\varphi^{j\Rightarrow i}(z,uq(\fbox{$j$}))\\
    &=\psi_{\emptyset}^{(i)}(z,u)\frac{N^{1\Rightarrow i}(z,u)}{D^{1\Rightarrow i}(z,u)}\frac{N^{j\Rightarrow i}(z,uq(\fbox{$j$}))}{D^{j\Rightarrow i}(z,uq(\fbox{$j$}))}.
    \end{split}
\end{align}
This charge function must contain poles where the atoms can be added or removed. In this case, since we cannot remove the atoms placed at the origin, the pole of the origin must be cancelled out with the numerators. The poles must contain the removable pole at the origin at level 1, so $N^{1\Rightarrow i}(z,u)$ cannot cancel out the pole at $z=u$.
Because $N^{j\Rightarrow i}(z,uq(\fbox{$j$}))$ must cancel the pole at the origin, we need the following relation:
\begin{align}
    N^{j\Rightarrow i}(z,uq(\fbox{$j$}))\supset z-u\quad \forall{j}\in[i\rightarrow].
\end{align}
We can set 
\begin{align}
   N^{j\Rightarrow i}(z,u)=\prod_{I\in\{i\rightarrow j\}}(q_{I}^{1/2}z-q_{I}^{-1/2}u).
\end{align}\label{eq:chargenumerator}
From (\ref{eq:chargedenominator}) and (\ref{eq:chargenumerator}), we finally obtain the charge function 
\begin{align}
    \varphi^{j\Rightarrow i}(z,u)=\frac{\prod_{I\in\{i\rightarrow j\}}(q_{I}^{1/2}z-q_{I}^{-1/2}u)}{\prod_{I\in\{j\rightarrow i\}}(q_{I}^{-1/2}z-q_{I}^{1/2}u)}.
\end{align}
By doing the similar discussion for general levels, one will see that the color $j$ atom at position $q(\fbox{$j$})$ gives a contribution to the charge $i$ function with a factor
\begin{align}
    \varphi^{j\Rightarrow i}(z,uq(\fbox{$j$}))
\end{align}
and the total charge function of a general crystal configuration will become (\ref{eq:chargeansatz}). For the discussion of general levels, we note that not all atoms are possible to add to the crystal configuration because of the melting rule. In the quiver Yangian case, for the crystal configuration to obey the melting rule, we need to impose the loop condition on the parameters:
\begin{align}
    \sum_{I\in\text{loop }L}h_{I}=0.
\end{align}
In the quantum toroidal case, this will be 
\begin{align}
    \prod_{I\in\text{loop }L}q_{I}=1.\label{eq:crystal_loopcond}
\end{align}
(See section 6.4 of \cite{Li:2020rij} for a nice discussion of the quiver Yangian case.)

\subsection{Bootstrapping the algebra when \texorpdfstring{$C=1$}{C=1}}\label{sec:EE,EFrelation}
Let us see that the action of the generators on the three-dimensional crystal indeed gives a representation of the algebra (\ref{eq:defofQuiverAlgebra}) when one of the central charges is $C=1$. We will only derive the EE relation and the EF relation. Other relations (KK, KE) can be discussed similarly and are in appendix \ref{sec:appendix_bootstrapp}.
We summarize here the results we obtained up to the previous subsection.
\begin{align}
\begin{split}
&K_{i}^{\pm}(z)\ket{\Lambda}=\left[\Psi_{\Lambda}^{(i)}(z,u)\right]_{\pm}\ket{\Lambda},\\
&\Psi_{\Lambda}^{(i)}(z,u)=\psi_{\emptyset}^{(i)}(z,u)\prod_{j\in Q_{0}}\prod_{\fbox{$j$}\in \Lambda}\varphi^{j\Rightarrow i}(z,uq(\fbox{$j$})),\\
&\varphi^{j\Rightarrow i}(z,u)=\frac{\prod_{I\in\{i\rightarrow j\}}(q_{I}^{1/2}z-q_{I}^{-1/2}u)}{\prod_{I\in\{j\rightarrow i\}}(q_{I}^{-1/2}z-q_{I}^{1/2}u)},\\
&E_{i}(z)\ket{\Lambda}=\sum_{\fbox{$i$}\in\text{Add}(\Lambda)}\pm\sqrt{p^{(i)}\underset{x=uq(\fbox{$i$})}{\Res}\Psi_{\Lambda}^{(i)}(x,u)}\delta\left(\frac{z}{uq(\fbox{$i$})}\right)\ket{\Lambda+\fbox{$i$}},\\
&F_{i}(z)\ket{\Lambda}=\sum_{\fbox{$i$}\in\text{Rem}(\Lambda)}\pm\sqrt{q^{(i)}\underset{y=uq(\fbox{$i$})}{\Res}\Psi_{\Lambda}^{(i)}(y,u)}\delta\left(\frac{z}{uq(\fbox{$i$})}\right)\ket{\Lambda-\fbox{$i$}},
\end{split}\label{eq:summaryofalgebraansatz}
\end{align}
where the parameters associated to the arrow of the quiver are subjected to the loop condition (\ref{eq:crystal_loopcond}). 

\subsubsection{EF relation}
Next, let us consider the commutation relations of $E_{i}(z)$ and $F_{j}(w)$. The relation $[E_{i}(z),F_{j}(w)]$ becomes either commutation relation or anti-commutation relation depending on the statistics of the operators.
\begin{align}
    \begin{split}
        &E_{i}(z)F_{j}(w)\ket{\Lambda}\\
        =&\sum_{\fbox{$j$}\in\text{Rem}(\Lambda)}\sum_{\fbox{$i$}\in \text{Add}(\Lambda-\fbox{$j$})}E^{(i)}(\Lambda-\fbox{$j$}\rightarrow \Lambda-\fbox{$j$}+\fbox{$i$})F^{(j)}(\Lambda\rightarrow \Lambda-\fbox{$j$})\\
        &\times\delta\left(\frac{z}{uq(\fbox{$i$})}\right)\delta\left(\frac{w}{uq(\fbox{$j$})}\right)\ket{\Lambda-\fbox{$j$}+\fbox{$i$}}
    \end{split}\label{eq:EFtotal}
\end{align}
\begin{align}
    \begin{split}
        F_{j}(w)&E_{i}(z)\ket{\Lambda}\\
        =&\sum_{\fbox{$i$}\in \text{Add}(\Lambda)}\sum_{\fbox{$j$}\in\text{Rem}(\Lambda+\fbox{$i$})}F^{(j)}(\Lambda+\fbox{$i$}\rightarrow \Lambda+\fbox{$i$}-\fbox{$j$})E^{(i)}(\Lambda\rightarrow \Lambda+\fbox{$i$})\\
        &\times\delta\left(\frac{z}{uq(\fbox{$j$})}\right)\delta\left(\frac{w}{uq(\fbox{$i$})}\right)\ket{\Lambda+\fbox{$i$}-\fbox{$j$}}.
    \end{split}\label{eq:FEtotal}
\end{align}
There are three situations in the sum above:
\begin{enumerate}
    \item $i=j$ and the atom added by $E_{i}(z)$ and the atom removed by $F_{j}(w)$ are the same.
    \item $i=j$ but the atom added by $E_{i}(z)$ and the atom removed by $F_{j}(w)$ are different.
    \item $i\neq j$, which means the atom added and removed are different.
 \end{enumerate}
Let us only consider the first situation. Only this situation gives the nontrivial terms on the right-hand side. Other situations will cancel out with each other and disappear after choosing the sign factors properly. See \cite{Li:2020rij} for the discussions of how to determine the sign factors.

In this case, in the sum of (\ref{eq:EFtotal}) and (\ref{eq:FEtotal}) we have the following terms:
\begin{align}
    \begin{split}
        \sum_{\fbox{$i$}\in\text{Rem}(\Lambda)}E^{(i)}(\Lambda-\fbox{$i$}\rightarrow \Lambda)F^{(i)}(\Lambda\rightarrow \Lambda-\fbox{$i$})\delta\left(\frac{z}{uq(\fbox{$i$})}\right)\delta\left(\frac{w}{uq(\fbox{$i$})}\right)\ket{\Lambda}\\
        -(-1)^{|i|}\sum_{\fbox{$i$}\in\text{Add}(\Lambda)}F^{(i)}(\Lambda+\fbox{$i$}\rightarrow \Lambda)E^{(i)}(\Lambda\rightarrow \Lambda+\fbox{$i$})\delta\left(\frac{z}{uq(\fbox{$i$})}\right)\delta\left(\frac{w}{uq(\fbox{$i$})}\right)\ket{\Lambda}.
    \end{split}\label{eq:EFsituation1}
\end{align}
Using 
\begin{align}
    \underset{y=uq(\fbox{$i$})}{\Res}\Psi_{\Lambda+\fbox{$i$}}^{(i)}(y,u)=\varphi^{i\Rightarrow i}(1,1) \underset{y=uq(\fbox{$i$})}{\Res}\Psi_{\Lambda}^{(i)}(y,u)
\end{align}
and setting 
\begin{align}
    \begin{split}
        &\sqrt{p^{(i)}q^{(i)}\varphi^{i\Rightarrow i} (1,1)}=1,\\
        &\epsilon(\Lambda-\fbox{$i$}\rightarrow \Lambda)\epsilon(\Lambda\rightarrow \Lambda-\fbox{$i$})=1,
    \end{split}\label{eq:EFconditionofsigns}
\end{align}
equation (\ref{eq:EFsituation1}) becomes
\begin{align}
    \begin{split}
        &\left(\sum_{\fbox{$i$}\in\text{Rem}(\Lambda)}+\sum_{\fbox{$i$}\in\text{Add}(\Lambda)}\right)\underset{y=uq(\fbox{$i$})}{\Res}\Psi^{(i)}_{\Lambda}(y,u)\delta\left(\frac{z}{uq(\fbox{$i$})}\right)\delta\left(\frac{w}{uq(\fbox{$i$})}\right)\ket{\Lambda}\\
        =&\delta\left(\frac{w}{z}\right)(K_{i}^{+}(z)-K_{i}^{-}(w))\ket{\Lambda}.
    \end{split}\label{eq:EFsituation1final}
\end{align}
At the last equation, we used (\ref{eq:residue}) and the property that $\Psi_{\Lambda}^{(i)}(z,u)$ has poles where the atoms can be added or removed. 

We also note 
\begin{align}
    \varphi^{i\Rightarrow i}(1,1)=(-1)^{\# \text{ of self-loops of $i$}}
\end{align}
and  that we need to impose the following condition
\begin{align}
    \varphi^{i\Rightarrow i }(1,1)=-(-1)^{|i|}\label{eq:statisticsofoperators}
\end{align}
to change (\ref{eq:EFsituation1}) to (\ref{eq:EFsituation1final}).
This condition shows that the statistics of the operators are related to the number of self-loops of the vertex in the quiver diagram. From this condition and (\ref{eq:EFconditionofsigns}), we also see 
\begin{align}
    p^{(i)}q^{(i)}=\varphi^{i\Rightarrow i}(1,1)=-(-1)^{|i|}.
\end{align}
 We can set $q^{(i)}=1$ without losing generality.

The EF relation is 
\begin{align}
     [E_{i}(z),F_{j}(w)]=\delta_{i,j}&\left(\delta\left(\frac{w}{z}\right)K_{i}^{+}(z)-\delta\left(\frac{z}{w}\right)K_{i}^{-}(w)\right).\label{eq:EFrelation}
\end{align}

\subsubsection{EE relation}\label{sec:EErelation}
We consider the action of $E_{i}(z)$ and $E_{j}(w)$ on a crystal configuration $\ket{\Lambda}$. Acting $E_{j}(w)$ first and $E_{i}(z)$ second, we get
\begin{align}
    \begin{split}
    &E_{i}(z)E_{j}(w)\ket{\Lambda}\\
    =&\sum_{\fbox{$j$}\in\text{Add}(\Lambda)}\sum_{\fbox{$i$}\in\text{Add}(\Lambda+\fbox{$j$})}\delta\left(\frac{z}{uq(\fbox{$i$})}\right)\delta\left(\frac{w}{uq(\fbox{$j$})}\right)\\
    &\quad\quad\times E^{(i)}(\Lambda+\fbox{$j$}\rightarrow \Lambda+\fbox{$j$}+\fbox{$i$})E^{(j)}(\Lambda\rightarrow \Lambda+\fbox{$j$})\ket{\Lambda+\fbox{$j$}+\fbox{$i$}}.
    \end{split}
\end{align}
Acting these operators in the opposite order, we obtain
\begin{align}
    \begin{split}
    &E_{j}(w)E_{i}(z)\ket{\Lambda}\\
    =&\sum_{\fbox{$i$}\in\text{Add}(\Lambda)}\sum_{\fbox{$j$}\in\text{Add}(\Lambda+\fbox{$i$})}\delta\left(\frac{z}{uq(\fbox{$i$})}\right)\delta\left(\frac{w}{uq(\fbox{$j$})}\right)\\
    &\quad\quad\times E^{(i)}(\Lambda+\fbox{$i$}\rightarrow \Lambda+\fbox{$i$}+\fbox{$j$})E^{(j)}(\Lambda\rightarrow \Lambda+\fbox{$i$})\ket{\Lambda+\fbox{$i$}+\fbox{$j$}}.
    \end{split}
\end{align}
In the generic situation when $\fbox{$i$}$ and $\fbox{$j$}$ do not depend on each other, the ratio of the coefficient of each term is 
\begin{align}
    \begin{split}
        &\frac{E^{(i)}(\Lambda+\fbox{$j$}\rightarrow \Lambda+\fbox{$j$}+\fbox{$i$})E^{(j)}(\Lambda\rightarrow \Lambda+\fbox{$j$})}{E^{(i)}(\Lambda+\fbox{$i$}\rightarrow\Lambda+\fbox{$i$}+\fbox{$j$})E^{(j)}(\Lambda\rightarrow\Lambda+\fbox{$i$})}\\
        =&\frac{\epsilon(\Lambda+\fbox{$j$}\rightarrow \Lambda+\fbox{$j$}+\fbox{$i$})}{\epsilon(\Lambda+\fbox{$i$}\rightarrow\Lambda+\fbox{$i$}+\fbox{$j$})}\frac{\epsilon(\Lambda\rightarrow \Lambda+\fbox{$j$})}{\epsilon(\Lambda\rightarrow\Lambda+\fbox{$i$})}\sqrt{\frac{\underset{y=uq(\fbox{$i$})}{\Res}\Psi^{(i)}_{\Lambda+\fbox{$j$}}(y,u)}{\underset{y=uq(\fbox{$i$})}{\Res}\Psi^{(i)}_{\Lambda}(y,u)}\frac{\underset{y=uq(\fbox{$j$})}{\Res}\Psi_{\Lambda}^{(j)}(y,u)}{\underset{y=uq(\fbox{$j$})}{\Res}\Psi_{\Lambda+\fbox{$i$}}^{(j)}(y,u)}}\\
        =&(-1)^{|i||j|}\sqrt{\frac{\underset{y=uq(\fbox{$i$})}{\Res}\varphi^{j\Rightarrow i}(y,uq(\fbox{$j$}))}{\underset{y=uq(\fbox{$j$})}{\Res}\varphi^{i\Rightarrow j}(y,uq(\fbox{$i$}))}}=(-1)^{|i||j|}\varphi^{j\Rightarrow i}(uq(\fbox{$i$}),uq(\fbox{$j$})),
    \end{split}\label{eq:EEassoc}
\end{align}
where in the last line we set 
\begin{align}
    \frac{\epsilon(\Lambda+\fbox{$j$}\rightarrow \Lambda+\fbox{$j$}+\fbox{$i$})}{\epsilon(\Lambda+\fbox{$i$}\rightarrow\Lambda+\fbox{$i$}+\fbox{$j$})}\frac{\epsilon(\Lambda\rightarrow \Lambda+\fbox{$j$})}{\epsilon(\Lambda\rightarrow\Lambda+\fbox{$i$})}=(-1)^{|i||j|}\label{eq:EEsign}
\end{align}
and used (\ref{eq:assoc}). Because of the delta functions $\delta\left(\frac{z}{uq(\fbox{$i$})}\right)$ and $\delta\left(\frac{w}{uq(\fbox{$j$})}\right)$, we obtain the following EE relation:
\begin{align}
    E_{i}(z)E_{j}(w)=(-1)^{|i||j|}\varphi^{j\Rightarrow i}(z,w)E_{j}(w)E_{i}(z).\label{eq:EErelation}
\end{align}
 Other defining relations can be derived similarly and the result is
\begin{align}
    \begin{split}
    K_{i}K_{i}^{-1}&=K_{i}^{-1}K_{i}=1,\\
     K^{\pm}_{i}(z)K^{\pm}_{j}(w)&=K^{\pm}_{j}(w)K^{\pm}_{i}(z),\\
    K^{\pm}_{i}(z)E_{j}(w)&=\varphi^{j\Rightarrow i}(z,w)E_{j}(w)K^{\pm}_{i}(z),\\
    K^{\pm}_{i}(z)F_{j}(w)&=\varphi^{j\Rightarrow i}(z,w)^{-1}F_{j}(w)K_{i}^{\pm}(z),\\
    [E_{i}(z),F_{j}(w)]=\delta_{i,j}&\left(\delta\left(\frac{w}{z}\right)K_{i}^{+}(z)-\delta\left(\frac{z}{w}\right)K_{i}^{-}(w)\right),\\
    E_{i}(z)E_{j}(w)&=(-1)^{|i||j|}\varphi^{j\Rightarrow i}(z,w)E_{j}(w)E_{i}(z),\\
    F_{i}(z)F_{j}(w)&=(-1)^{|i||j|}\varphi^{j\Rightarrow i}(z,w)^{-1}F_{j}(w)F_{i}(z).
    \end{split}\label{eq:algebraverticalrepdef}
\end{align}
These are indeed the defining relations of the quiver quantum toroidal algebra in (\ref{eq:defofQuiverAlgebra}) after setting $C=1$.

\subsection{Some issues on the asymmetric quiver} \label{sec:cpt4cycle}
  In this section, we summarize where we need to assume that the quiver is symmetric. We note that the asymmetric quiver is associated with the Calabi-Yau manifold with the compact 4-cycle. We need to modify our proposal to treat these general cases, which we leave for future work.
 \begin{itemize}
     \item Modes of $K_{i}^{\pm}(z)$ in (\ref{eq:QQTAgenerator}).\\
     We defined the mode expansions starting from $z^{\mp r}(r\geq 0)$. This is possible only for the symmetric case, which means $|i\rightarrow j|=|j\rightarrow i|$. To be concrete, we consider the KE relation of (\ref{eq:defofQuiverAlgebra}):
     \begin{align}
         K_{i}^{+}(z)E_{j}(w)=\varphi^{j\Rightarrow i}(z,w)E_{j}(w)K_{i}^{+}(z),
     \end{align}
     where we set $C=1$ to make discussions simple.
     When $|i\rightarrow j|=|j\rightarrow i|$, since
     \begin{align}
         \varphi^{i\Rightarrow j}(z,w)=\varphi^{i\Rightarrow j}(1,w/z),
     \end{align}
     the degree of $z$ of both hand sides match. However, when there are compact 4-cycles the bond factor will factorize as 
     \begin{align}
         \varphi^{i\Rightarrow j}(z,w)=z^{|j\rightarrow i|-|i\rightarrow j|}\varphi^{i\Rightarrow j}(1,w/z).
     \end{align}
     For the degrees of both hand sides to match, we need to change the mode expansions of $K^{+}(z)$ to include $z^{r}(r>0)$. A similar discussion goes for $K^{-}(z)$ and we need to include $z^{-r}(r>0)$ modes this time. Thus, for the asymmetric case, we have to face the mode expansions of $K^{\pm}_{i}(z)$ in the form
     \begin{align}
         K_{i}^{\pm}(z)=K_{i}^{\pm}\exp\left(\pm\sum_{r=-\infty}^{\infty}H_{i,\pm r}z^{\mp r}\right).
     \end{align}
     It is already known that a similar modification is necessary for the quiver Yangian case.
     
     \item The associativity condition $\varphi^{i\Rightarrow j}(z,w)\varphi^{j\Rightarrow i}(w,z)=1$ in (\ref{eq:assoc}) and (\ref{eq:assoc_no4cycle}). This condition was also used in deriving the EE relation (\ref{eq:EEassoc}) in section \ref{sec:EErelation}.\\
     By direct computation, the product of the bond factors is
     \begin{align}
         \varphi^{i\Rightarrow j}(z,w)\varphi^{j\Rightarrow i}(w,z)=(-1)^{|j\rightarrow i|+|i\rightarrow j|}=(-1)^{|j\rightarrow i|-|i\rightarrow j|}.
     \end{align}
    For the generators to have the associativity condition, we need the condition 
    \begin{align}
    (-1)^{|j\rightarrow i|+|i\rightarrow j|}=1.
    \end{align}
  
For the asymmetric case, this condition itself is not trivial, so we need some kind of modifications in the definition of the algebra\footnote{We note that such sign factors only affect a subclass of asymmetric quivers because if all $|i-j|-|j-i|$ are even, the sign factors vanish. The same issue is discussed on page 18 of the v3 of \cite{Li:2020rij}, where they address this by modifying the bond factors.}. We consider the quiver associated with the $K_{\mathbb{P}^{2}}$ geometry as an example. The toric diagram and quiver diagram are illustrated in Figure \ref{fig:KP2}. There are three vertices in the quiver diagram, and each of the vertices is connected to the other two vertices. The number of the arrows between these two vertices is not symmetric. This shows that $|1\rightarrow 2|=3$ and $|2\rightarrow 1|=0$, which implies $(-1)^{|1\rightarrow 2|+|2\rightarrow 1|}=-1\neq1$.  
     \begin{figure}[H]
         \centering
         \includegraphics[width=9cm]{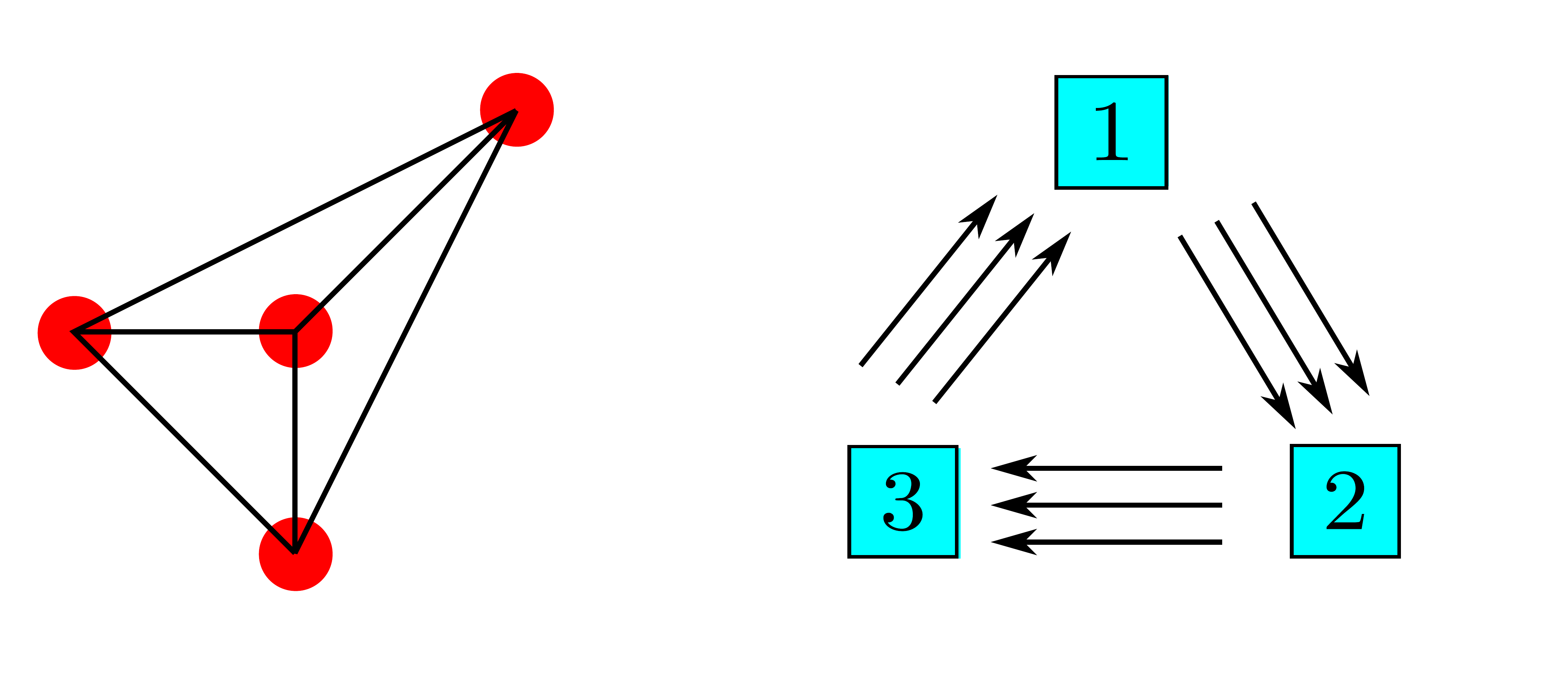}
         \caption{Toric diagram (left) and quiver diagram (right) of $K_{\mathbb{P}^{2}}$ geometry. The number of arrows between two vertices of the quiver diagram is not symmetric. }
         \label{fig:KP2}
     \end{figure}
     In deriving the EE relation from the action on the BPS crystal, we also need to assume the quiver is symmetric. For the asymmetric case, 
     \begin{align}
         \sqrt{\frac{\underset{y=uq(\fbox{$i$})}{\Res}\varphi^{j\Rightarrow i}(y,uq(\fbox{$j$}))}{\underset{y=uq(\fbox{$j$})}{\Res}\varphi^{i\Rightarrow j}(y,uq(\fbox{$i$}))}}=(-1)^{\frac{1}{2}(|i\rightarrow j|+|j\rightarrow i|)}\varphi^{j\Rightarrow i}(uq(\fbox{$i$}),uq(\fbox{$j$}))\,.
     \end{align}
     which implies the last line of (\ref{eq:EEassoc}) should be modified. We may absorb these extra signs by introducing extra $(-1)$ factors to the bond factors, which remains as a possibility. For now, we do not know how to determine this factor.\footnote{After submitting the first version of this paper to arXiv, a paper \cite{Galakhov:2021vbo} giving a sign choice appeared. See it for more discussions. See also the footnote on page 18 in v3 of \cite{Li:2020rij}. }
     
     \item  The Hopf superalgebra structure in section \ref{sec:Hopf_structure}.\\
     We focus on the definition of the coproduct structure. We assume that the algebra remains the same even when the quiver diagram is asymmetric, and the bond factors do not have the same number of zeros and poles. We assume that the coproduct formula is the same as in (\ref{eq:coproduct}). Acting the coproduct on the left-hand side of the $K^{-}E$ relation, we obtain 
     \begin{align}
     \begin{split}
    &\Delta(K_{i}^{-}(Cz)E_{j}(w))\\
    =&\varphi^{j\Rightarrow i}(z,w)E_{j}(w)K_{i}^{-}(C_{1}z)\otimes K_{i}^{-}(C_{1}C_{2}z)\\
    &\quad +\textcolor{blue}{\varphi^{j\Rightarrow i}(C_{1}z,C_{1}w)}K_{j}^{-}(C_{1}w)K_{i}^{-}(C_{1}z)\otimes E_{j}(C_{1}w)K_{i}^{-}(C_{1}C_{2}z)\\
    =&\varphi^{j\Rightarrow i}(z,w)\left\{E_{j}(w)K_{i}^{-}(C_{1}z)\otimes K_{i}^{-}(C_{1}C_{2}z)\right.\\
    &\left.\quad+\textcolor{blue}{C_{1}^{|i\rightarrow j|-|j\rightarrow i|}}K_{j}^{-}(C_{1}w)K_{i}^{-}(C_{1}z)\otimes E_{j}(C_{1}w)K_{i}^{-}(C_{1}C_{2}z)\right\},
    \end{split}
\end{align} 
     where in the last line we used (\ref{eq:no4cycle_bondfactor_symmetry}). Thus, for the asymmetric case, the algebra does not have the coproduct structure as it is. Similar discussions can be done for other defining relations and for the antipode, and we will see that the Hopf superalgebra structure is not well-defined anymore. 
     One way to resolve this inconsistency is to modify the defining relations of $K^{\pm}K^{\pm}$ as
     \begin{align}
          K_{i}^{\pm}(z)K_{j}^{\pm}(w)&=C^{\pm(|i\rightarrow j|-|j\rightarrow i|)}K_{j}^{\pm}(w)K_{i}^{\pm}(z).
     \end{align}
    One can show that after this modification, the formal Hopf superalgebra structure is recovered. This defining relation only appears in the case when we consider asymmetric quivers and representations with nontrivial central charge $C\neq 1$.
    \item The rescaling symmetry (\ref{eq:rescalesymm}) and vertex condition (\ref{eq:vertexconstraint}).\\
    In the equations (\ref{eq:vertexconstr_cpt4cond_1}), (\ref{eq:vertexconstr_cpt4cond_2}), and (\ref{eq:vertexconstr_cpt4cond_3}), we used the rescaling symmetry in (\ref{eq:rescalesymm}), which is true when there are no compact 4-cycles. The discussion there might be modified if we use a different convention of the bond factor.\footnote{This problem seems to be resolved in \cite{Galakhov:2021vbo}, so see it for more discussions.} 
   
\end{itemize}

\section{Example: \texorpdfstring{$\mathbb{C}^{3}/(\mathbb{Z}_{2}\times\mathbb{Z}_{2})$}{C3Z2Z2} and quantum toroidal \texorpdfstring{$D(2,1;\alpha)$}{D21}}\label{sec:Example}
We give a nontrivial example of the quiver quantum toroidal algebra, which is associated with the abelian orbifold $\mathbb{C}^{3}/(\mathbb{Z}_{2}\times\mathbb{Z}_{2})$. The toric diagram, dual web diagram, and periodic quiver are shown in Figure \ref{fig:D(2,1)toric_periodicquiver}. The quiver diagram derived from the periodic quiver is shown in Figure \ref{fig:D(2,1)quiver}, which is identical to the Dynkin diagram of the affine superalgebra $\hat{D}(2,1;\alpha)$ in Figure \ref{fig:D(2,1)root}. We suppose that this is the quantum toroidal $D(2,1;\alpha)$, which is yet to be studied in detail. See \cite{Feigin_2019,feigin2021combinatorics} for recent developments. For the Drinfeld second realization of the quantum affine superalgebras of $D(2,1;\alpha)$, see \cite{heckenberger2008drinfeld}.
\begin{figure}
    \begin{tabular}{cc}
      \begin{minipage}{0.45\hsize}
        \centering
       \includegraphics[width=7cm]{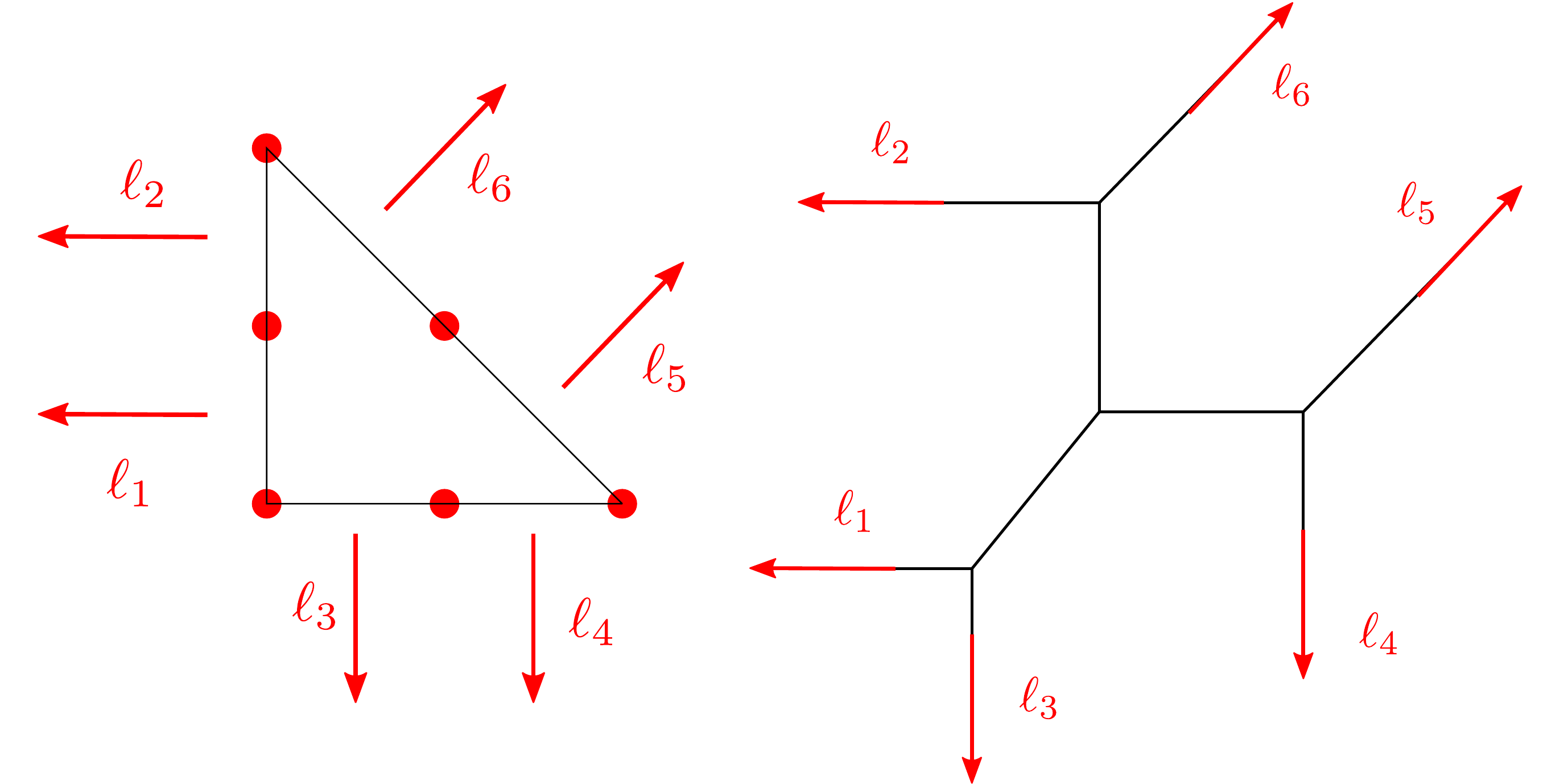}
        \subcaption{Toric diagram and dual web diagram.}\label{fig:D(2,1)toric_web}
      \end{minipage}&\hfill
      \begin{minipage}{0.45\hsize}
        \centering
     \includegraphics[width=7cm]{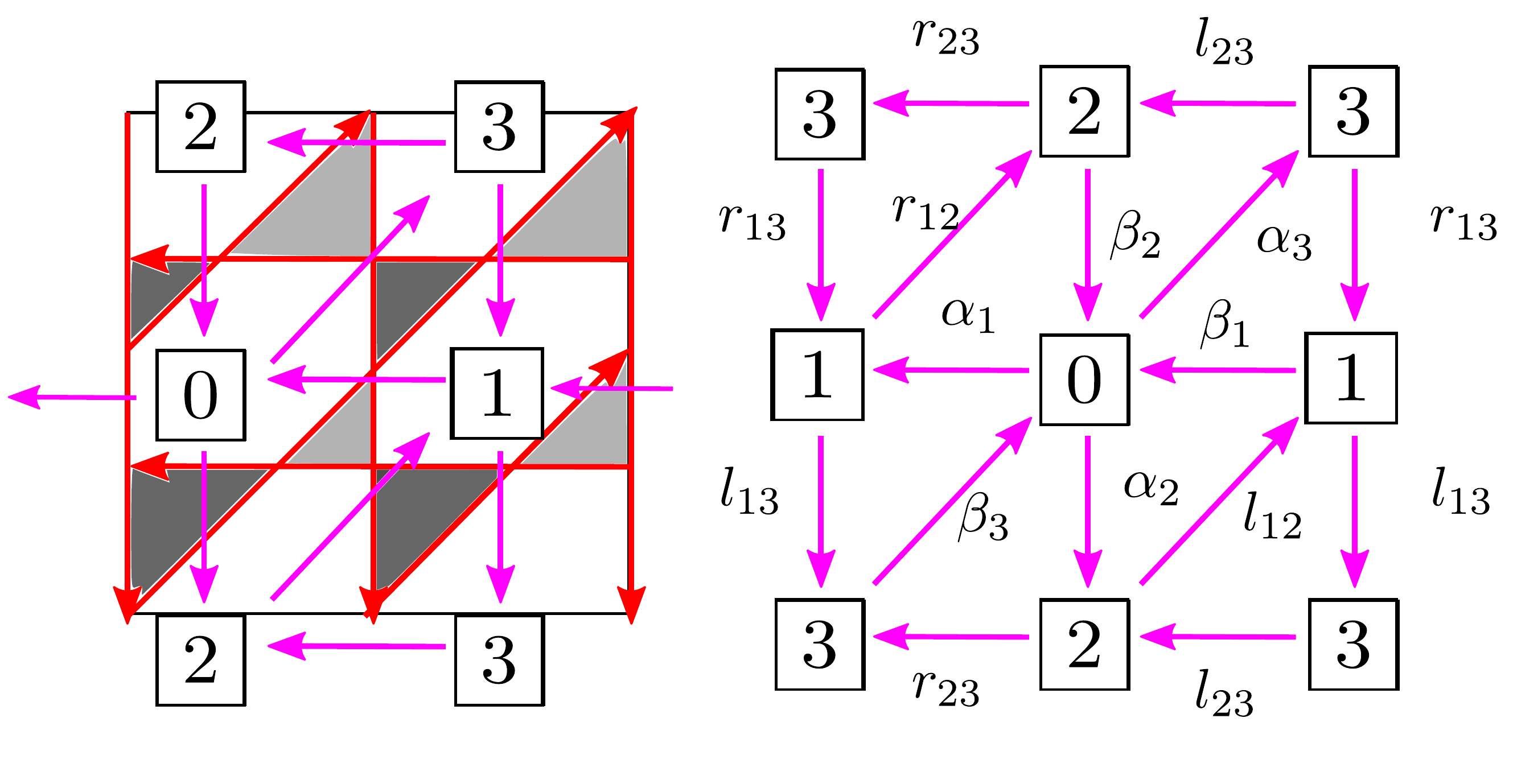}
       \subcaption{Periodic quiver.}\label{fig:D(2,1)periodic_quiver}
      \end{minipage}
    \end{tabular}
\caption{Toric diagram, dual web diagram, and periodic quiver of $\mathbb{C}^{3}/(\mathbb{Z}_{2}\times\mathbb{Z}_{2})$. The six lattice points of the toric diagram are denoted as $(0,0)$, $(1,0)$, $(2,0)$, $(0,1)$, $(0,2)$, and $(1,1)$.}\label{fig:D(2,1)toric_periodicquiver}
\end{figure}

\begin{figure}
    \begin{tabular}{cc}
      \begin{minipage}{0.45\hsize}
        \centering
      \includegraphics[width=5cm]{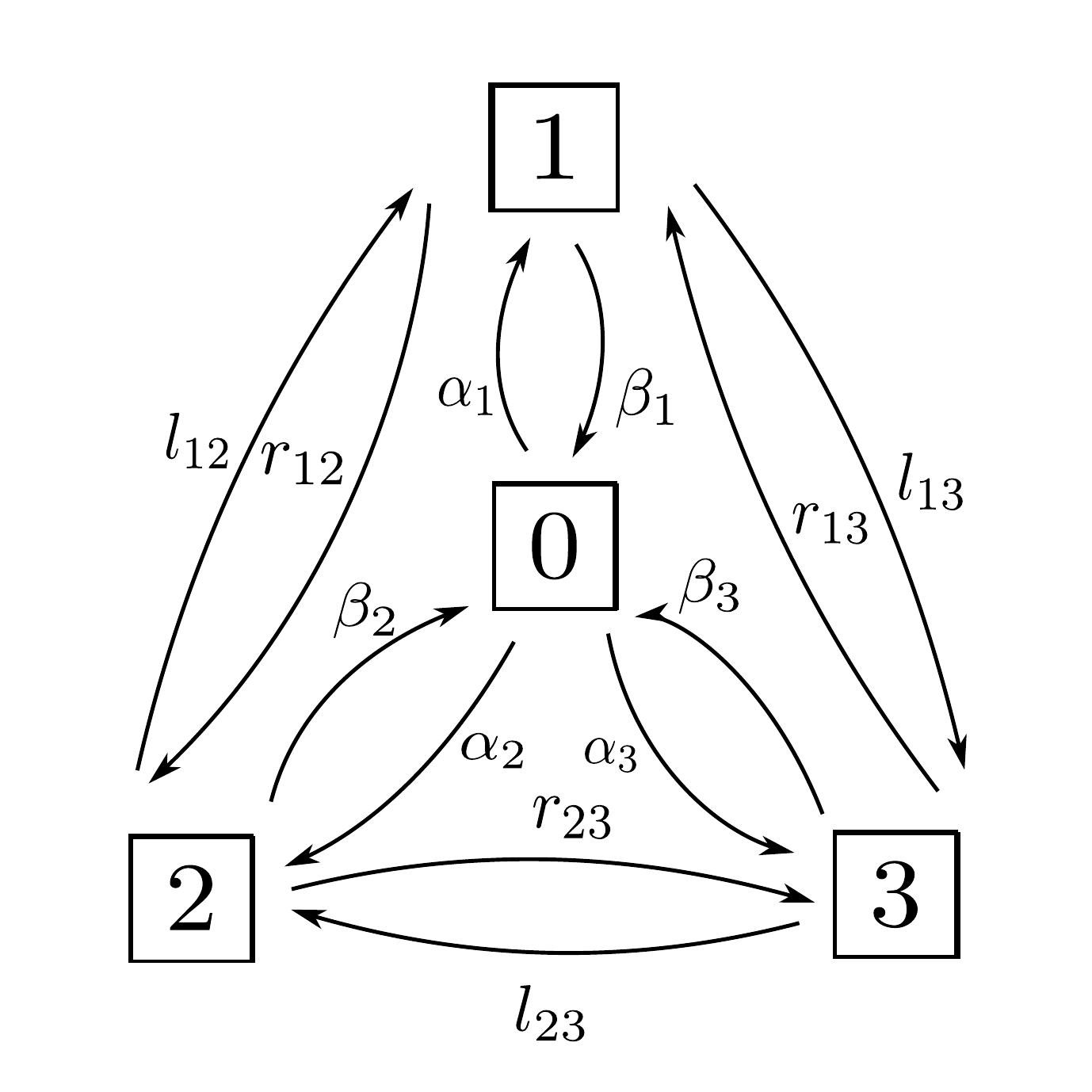}
        \subcaption{Quiver diagram.}\label{fig:D(2,1)quiver}
      \end{minipage}&\hfill
      \begin{minipage}{0.45\hsize}
        \centering
     \includegraphics[width=5cm]{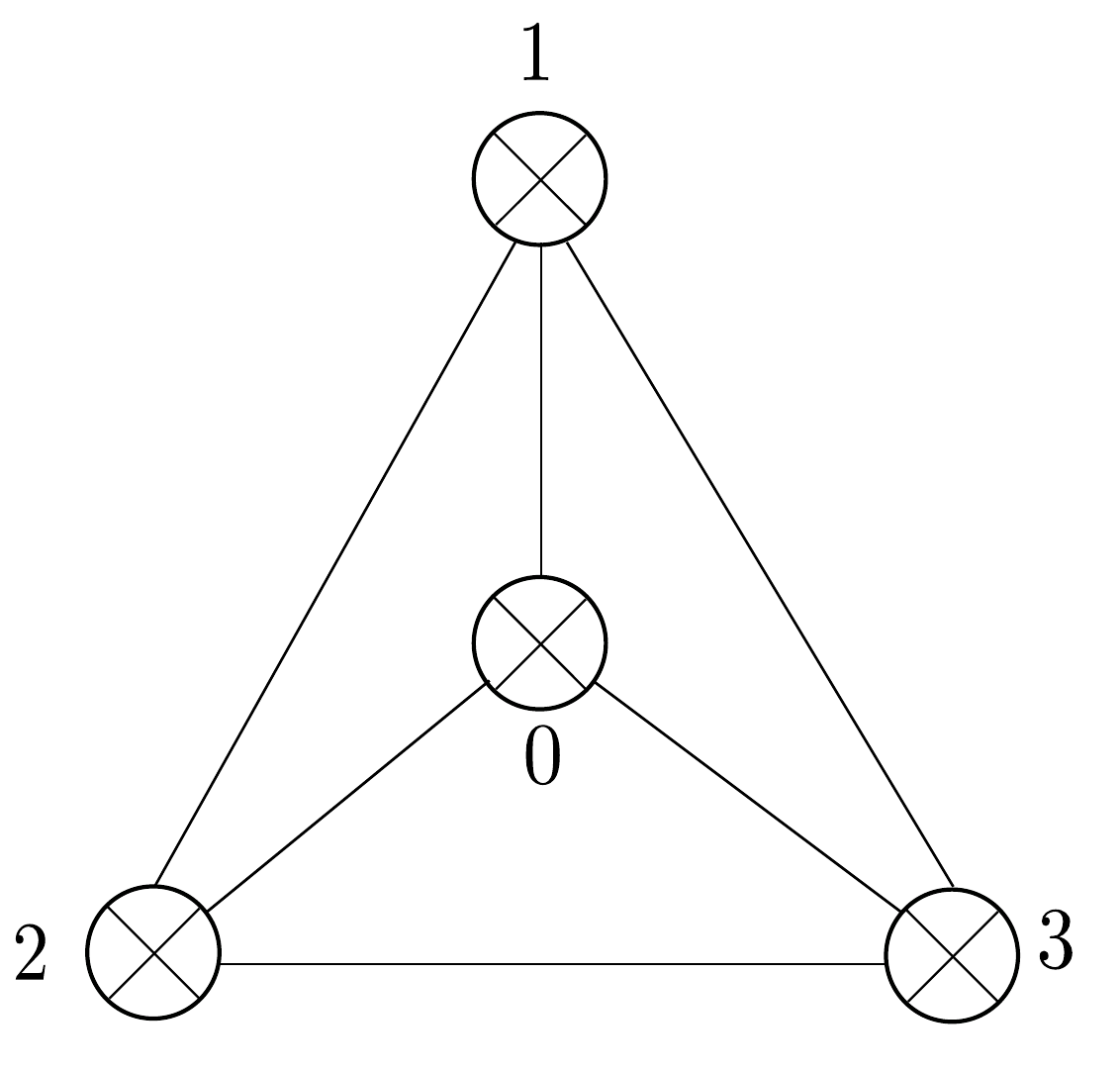}
       \subcaption{Dynkin diagram of $\hat{D}(2,1;\alpha)$.}\label{fig:D(2,1)root}
      \end{minipage}
    \end{tabular}
    \caption{Quiver diagram and Dynkin diagram of affine superalgebra $\hat{D}(2,1;\alpha)$. We choose all of the simple roots to be fermionic. }
\end{figure}

We first note that all the generators are fermionic since there are no loops for each vertex. 
Let us derive the bond factors of the algebra. We have 12 parameters which are assigned as Figure \ref{fig:D(2,1)periodic_quiver} and Figure \ref{fig:D(2,1)quiver}. The loop constraints are 
\begin{align}
\begin{split}
    r_{13}r_{12}r_{23}=1,\quad \alpha_{1}r_{12}\beta_{2}=1,\quad \beta_{2}\alpha_{3}l_{23}=1,\quad \alpha_{3}r_{13}\beta_{1}=1,\\
    \alpha_{1}l_{13}\beta_{3}=1,\quad \alpha_{2}r_{23}\beta_{3}=1,\quad \alpha_{2}l_{12}\beta_{1}=1,\quad l_{23}l_{12}l_{13}=1.
    \end{split}
\end{align}
We have 8 constraints and 7 of them are independent, so we get 5 parameters after imposing these conditions. The vertex constraints are
\begin{align}
\begin{split}
    \alpha_{1}\alpha_{2}\alpha_{3}=\beta_{1}\beta_{2}\beta_{3},\quad \alpha_{1}l_{12}r_{13}=l_{13}r_{12}\beta_{1},\\
    l_{12}\beta_{2}r_{23}=r_{12}\alpha_{2}l_{23},\quad \alpha_{3}r_{23}l_{13}=r_{13}\beta_{3}l_{23}
    \end{split}
\end{align}
and 3 of them are independent. After imposing all of the constraints we get two independent parameters:
\begin{align}
\begin{split}
\alpha_{1}=\beta_{1}=l_{23}=r_{23}=q_{1},\\
\alpha_{2}=\beta_{2}=l_{13}=r_{13}=q_{2},\\
\alpha_{3}=\beta_{3}=l_{12}=r_{12}=q_{3},
\end{split}
\end{align}
with the condition $q_{1}q_{2}q_{3}=1$.
Using these we obtain the bond factors:
\begin{align}
    \varphi^{i\Rightarrow j}(z,w)=\frac{\phi(q_{ij};z,w)}{\phi(q_{ij}^{-1};z,w)},\label{eq:D(2,1)bondfactors}
\end{align}
where
\begin{align*}
    q_{ij}=q_{ji}=\begin{cases} 
    q_{1}\quad(i,j)=(0,1),(2,3)\\
    q_{2}\quad(i,j)=(0,2),(1,3)\\
    q_{3}\quad(i,j)=(0,3),(1,2)
    \end{cases}\,.
\end{align*}

\begin{figure}[H]
       \centering
     \includegraphics[width=6cm]{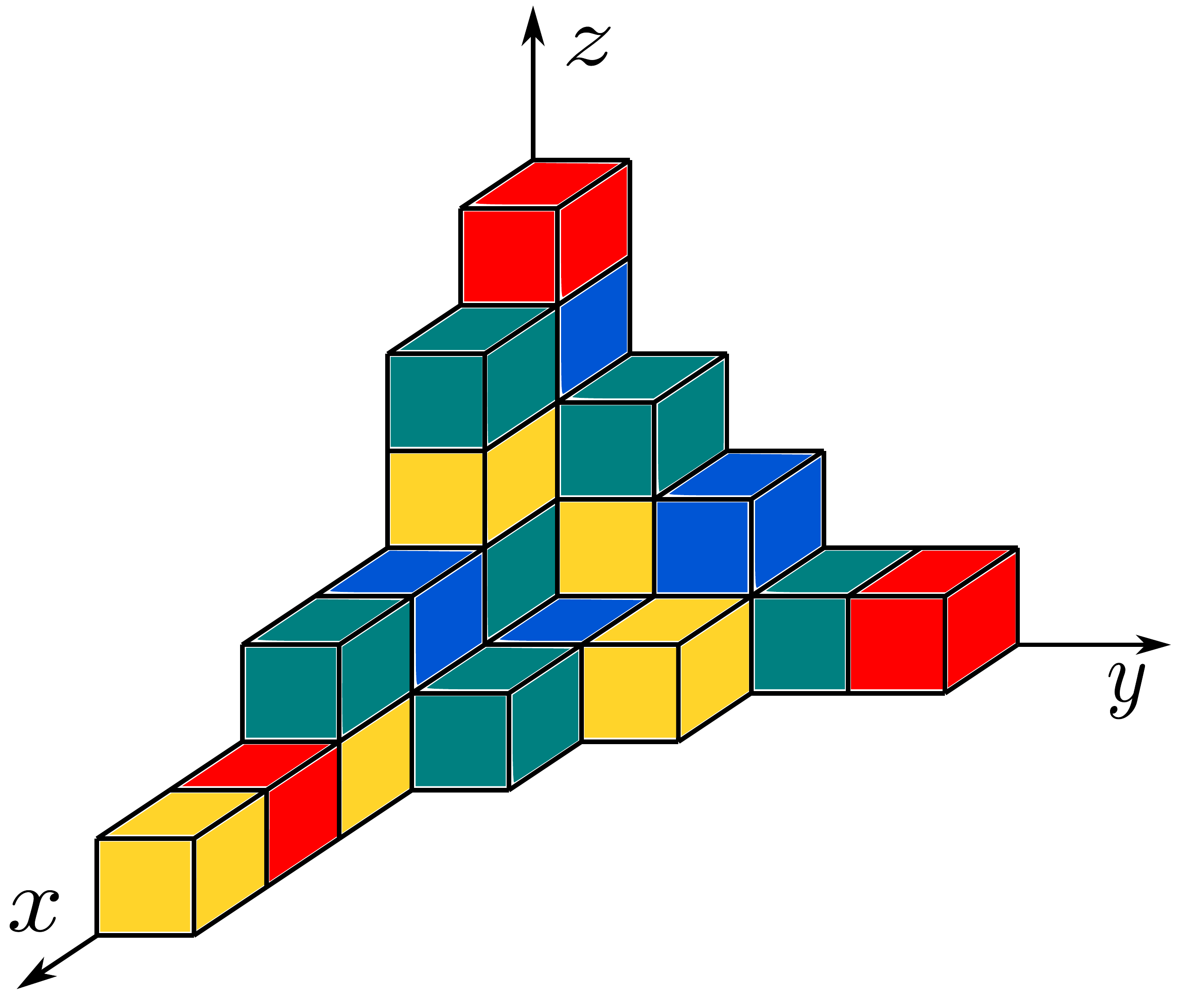}
\caption{Quiver diagram and three-dimensional crystal of $\mathbb{C}^{3}/(\mathbb{Z}_{2}\times\mathbb{Z}_{2})$. The three-dimensional crystal is a plane partition, which is the same as the quantum toroidal $\mathfrak{gl}_{1}$, but the coloring is different. There are four colors: red, blue, yellow, green. Each of them corresponds to the four vertices of the quiver diagram. The origin box is red.} \label{fig:D(2,1)three-dimensionalcrystal}
\end{figure}

\begin{figure}[H]
    \begin{tabular}{cc}
      \begin{minipage}{0.45\hsize}
        \centering
      \includegraphics[width=5cm]{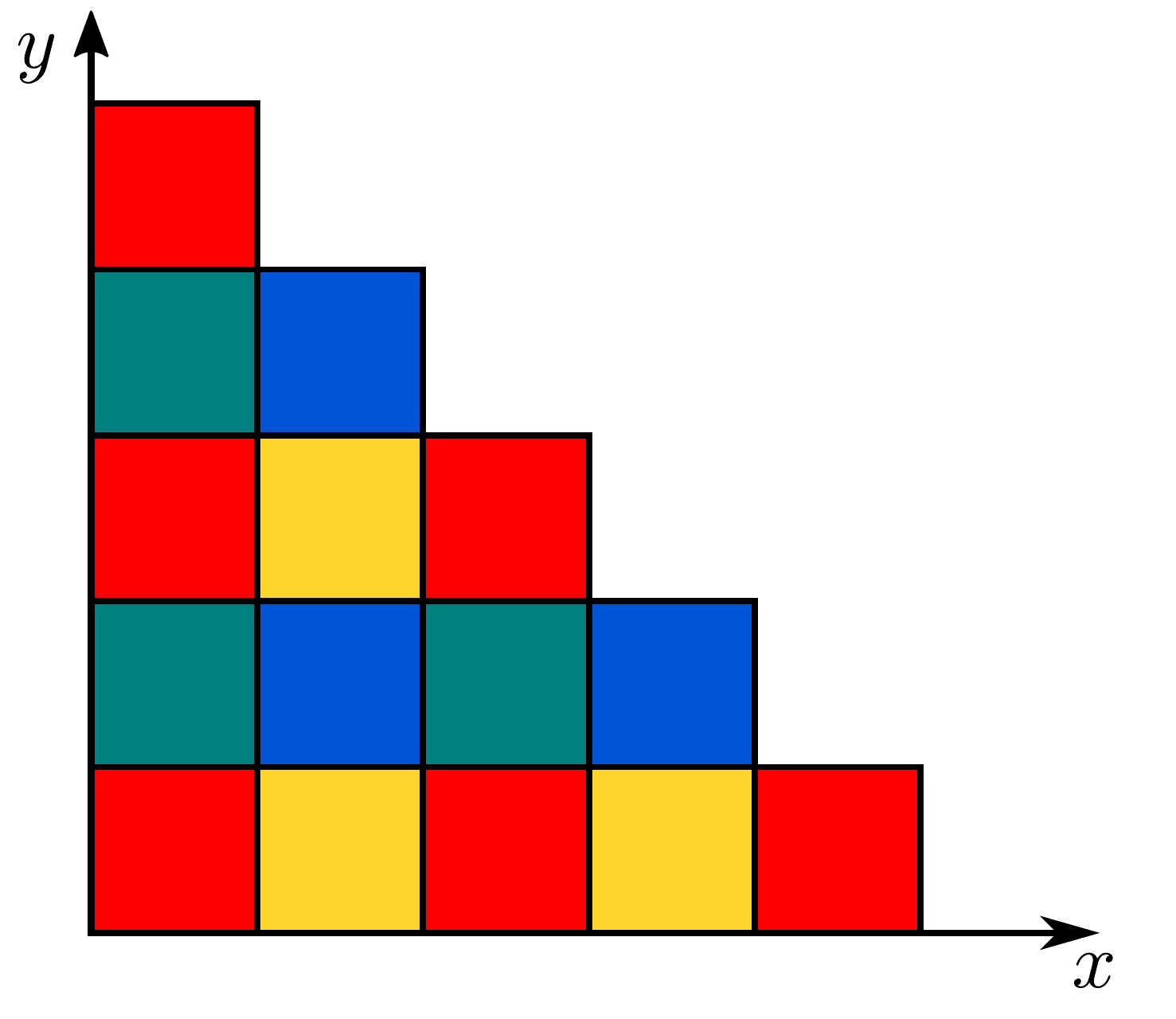}
        \subcaption{$k=1$}\label{fig:D(2,1)z=0}
      \end{minipage}&\hfill
      \begin{minipage}{0.45\hsize}
        \centering
     \includegraphics[width=5cm]{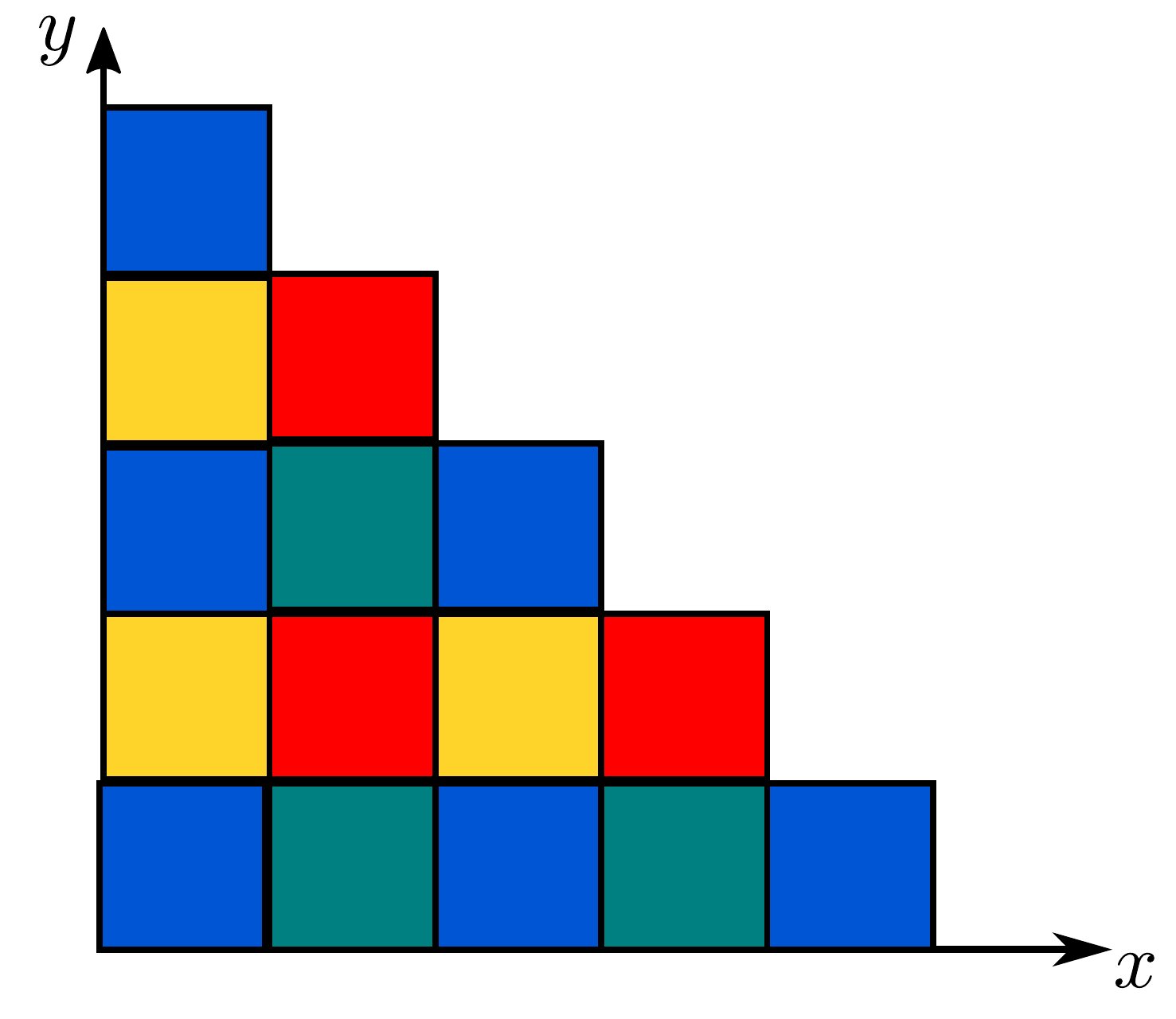}
       \subcaption{$k=2$}\label{fig:D(2,1)z=1}
      \end{minipage}
    \end{tabular}
    \caption{Slice of the three-dimensional crystal at $k=1$ and $k=2$. The coordinates of the boxes are assigned $(i,j,k)$ $(i,j,k\in\mathbb{Z}_{>0})$. (a) Patterns of coloring of boxes when $k=1$. (b) Patterns of coloring of boxes when $k=2$. }
\end{figure}

From the periodic quiver diagram in Figure \ref{fig:D(2,1)periodic_quiver}, we obtain the three-dimensional crystal in Figure \ref{fig:D(2,1)three-dimensionalcrystal}. While the shape of the crystal is the same as the plane partition representation for the quantum toroidal $\mathfrak{gl}_{1}$, we have four colors (red, blue, green, and yellow) to fill the boxes. Each box is stacked in such a way that no two adjacent boxes have the same color. For example, the red box has yellow boxes next to it in the $x$ direction, green boxes in the $y$ direction, and blue boxes in the $z$ direction (see Figure \ref{fig:D(2,1)three-dimensionalcrystal}).

 Let the three-dimensional coordinates of the boxes be $(i,j,k)$ $(i,j,k\in\mathbb{Z}_{>0})$. We note the origin is $(1,1,1)$ and the color is red. Using $q_{1},q_{2},q_{3}$, the coordinate is written as $q_{1}^{i-1}q_{2}^{j-1}q_{3}^{k-1}$. The $z=1$ plane has boxes colored as in Figure \ref{fig:D(2,1)z=0}. It is obvious that the color of the box is red when $i\equiv1(\hspace{-2mm}\mod2)$ and $j\equiv1$, yellow when $i\equiv0$ and $j\equiv1$, green when $i\equiv1$ and $j\equiv0$, and blue when $i\equiv0$ and $j\equiv0$. The color of the box in $(i,j,k)$ can be determined similarly, and it is red when $i-k\equiv0$ and $j-k\equiv0$, yellow when $i-k\equiv1$ and $j-k\equiv0$, green when $i-k\equiv0$ and $j-k\equiv1$, and blue when $i-k\equiv1$ and $j-k\equiv1$ (see Figure \ref{fig:D(2,1)z=1} for $k=2$). We note the equality is understood modulo 2.

As one can see, although the coloring pattern of the boxes is different from the Fock representations of the quantum toroidal $\mathfrak{gl}_{1}$ or the quantum toroidal $\mathfrak{gl}_{n}$, the shape itself is the same. Thus, one would like to ask whether it is a representation of the quantum toroidal algebra defined from the bond factors (\ref{eq:D(2,1)bondfactors}). Further studies will be done in \cite{Noshita2021}.

\section{Summary and Discussions}\label{sec:summary}
We defined the quiver quantum toroidal algebra associated with toric Calabi-Yau threefolds without compact 4-cycles. We introduced a central element $C$ and showed that the algebra is an associative Hopf superalgebra. When the central element $C$ is trivial ($C=1$), one of the representations is indeed the three-dimensional BPS crystal introduced in \cite{Ooguri_2009}, and the algebra can be bootstrapped following the strategy of \cite{Li:2020rij}. We leave general discussions for representations with nontrivial central charges of $C$  for future work.

As an example, we introduced a quiver quantum toroidal algebra associated with the orbifold $\mathbb{C}^{3}/(\mathbb{Z}_{2}\times\mathbb{Z}_{2})$. The quiver diagram is the same as the Dynkin diagram of the affine superalgebra $D(2,1;\alpha)$. We expect the quantum toroidal algebra associated with this affine superalgebra is the one we defined. The three-dimensional crystal of it is the plane partition, which is the same shape as the MacMahon representation of the quantum toroidal $\mathfrak{gl}_{1}$. Although the shape is the same, the coloring of the boxes is different. There are four colors, and each box is stacked in such a way that no two adjacent boxes have the same color. Since this representation is similar to the MacMahon representation of the quantum toroidal $\mathfrak{gl}_{1}$, we expect there are also analogs of ``Fock" representations and ``vector" representations of the quantum toroidal $\mathfrak{gl}_{1}$. The ``Fock" representations should be associated with the $(x,y)$, $(y,z)$, $(z,x)$ planes of the colored MacMahon representation, while the ``vector"  representations should be associated with the $x$, $y$, $z$ axes, as the toroidal $\mathfrak{gl}_{1}$ case. These properties are discussed in the companion paper \cite{Noshita2021}.

Although we focused on the case when there are no compact 4-cycles, the discussions should be extended to arbitrary toric Calabi-Yau threefolds including compact 4-cycles. Since the bond factors do not have the same number of zeros and poles anymore, we expect $K^{\pm}(z)$ should include all degrees of $z$, which is a similar situation to the quiver Yangian case \cite{Li:2020rij}. However, as mentioned in section \ref{sec:cpt4cycle}, we have used the no compact 4-cycle condition in various places. Thus, modifications of the algebra are necessary for generalizations. We hope to come back to this in the near future.
Finally, let us list down some possible directions we hope to clarify.
\begin{itemize}
    \item Horizontal representations ($C\neq 1$): As mentioned above, the central element $C$ we introduced is still a conjecture. Representations of $C\neq 1$ are expected to be related directly to the $q$ deformed version of the rational $\mathcal{W}$ algebras, which are associated with truncations of the three-dimensional BPS crystal representation (see section 7 of \cite{Li:2020rij}). To make it concrete, let us consider the $\mathbb{C}^{3}$-geometry case. The plane partition representation of the affine Yangian $\mathfrak{gl}_{1}$ has truncations, and it is understood as a ``pit" reduction \cite{bershtein2018plane}. The corresponding algebra is denoted as $Y_{L, M, N}$, where $(L, M, N)$ is the location of the pit, and called corner VOA (CVOA) \cite{Gaiotto:2017euk}. Free field realizations were derived in \cite{Litvinov_2016, Prochazka:2018tlo}. We can consider a $q$-deformation of the CVOA \cite{Harada_2021,bershtein2018plane, Kojima2019, Kojima2021, FHSSY:2010}, and it is obtained by taking tensor products of Fock representations. These Fock representations are associated with the divisors of $\mathbb{C}^{3}$ and have a nontrivial central charge of $C$. We expect this will be the same situation for general toric Calabi-Yau manifolds. Namely, we expect there are horizontal representations associated with the divisors of the toric Calabi-Yau, and by taking tensor products of them, we can obtain the corresponding $q$-deformed $\mathcal{W}$ algebra. 
    \item Serre relations, Miki automorphism: In this paper, we omit the discussions of Serre relations and Miki automorphism. Serre relations for $\mathfrak{gl}_{1}$, $\mathfrak{gl}_{n}$, and $\mathfrak{gl}_{m|n}$ are already known, but unknown for $D(2,1;\alpha)$. Similar discussions of \cite{Li:2020rij} might help solve this problem. We also expect there is Miki automorphism relating the vertical representations with the horizontal representations. 
    \item Generalizations to general orbifolds: A new quantum toroidal algebra associated with $\mathbb{C}\times\mathbb{C}^{2}/\mathbb{Z}_{p}$, where the action of $\mathbb{Z}_{p}$ is determined by two integers was introduced in \cite{Bourgine_2020}. We expect we can do a similar deformation of the quantum toroidal algebra of $\mathfrak{gl}_{m|n}$ and $D(2,1;\alpha)$. 
\end{itemize}

\acknowledgments 
The authors thank Koichi Harada and Yutaka Matsuo for useful discussions. 
GN is supported in part by FoPM, the University of Tokyo. AW is supported in part by JSPS fellowship, MEXT, and JSR Fellowship, the University of Tokyo.

\appendix
\section{Convention}\label{s:Notation}
We summarize the conventions and few residue formulas used in this paper. 
\begin{align}
	\begin{split}
		&\phi(a;z,w)\equiv a^{1/2}z-a^{-1/2}w,\\
		&\frac{\phi(a;z,w)}{\phi(b;z,w)}=\frac{a^{1/2}z-a^{-1/2}w}{b^{1/2}z-b^{-1/2}w},\\
		&\frac{\phi(a;z,pw)}{\phi(b;z,pw)}=\frac{\phi(ap^{-1};z,u)}{\phi(bp^{-1};z,u)},\\
		&\frac{\phi(a;pz,w)}{\phi(b;pz,w)}=\frac{\phi(ap;z,w)}{\phi(bp;z,w)}
	\end{split}
\end{align}
The formal expansion of the delta function is 
\begin{align}
    \delta(z)=\sum_{n\in \mathbb{Z}}z^{n}.
\end{align}
Two formal expansions $\left[\quad\right]_{\pm}$ are defined as 
\begin{align}
    \left[\frac{1}{\phi(a;z,w)}\right]_{+}\equiv \frac{1}{a^{1/2}z}\sum_{n\geq 0}\left(\frac{w}{az}\right)^{n},\\
    \left[\frac{1}{\phi(a;z,w)}\right]_{-}\equiv-\frac{1}{a^{-1/2}w}\sum_{n\geq 0}\left(\frac{az}{w}\right)^{n}.
\end{align}
Namely, $[\phi(a;z,w)]_{\pm}$ are formal expansions of $(\frac{z}{w})^{\mp1}$.

The residue formulas can be defined as the following:
\begin{align}
	\begin{split}
		\left[\frac{\phi(p;z,w)}{\phi(q;z,w)}\right]_{+}-\left[\frac{\phi(p;z,w)}{\phi(q;z,w)}\right]_{-}&=\phi(pq^{-1};1,1)\delta\left(\frac{z}{wq^{-1}}\right)\\
		&=\Res_{z=wq^{-1}}\frac{\phi(p;z,w)}{\phi(q;z,w)}\delta\left(\frac{z}{q^{-1}w}\right).
	\end{split}
\end{align}
We note 
\begin{align}
	\left[\phi(p;z,w)\right]_{+}-\left[\phi(p;z,w)\right]_{-}=0.
\end{align}
Other useful formulas are 
\begin{align}
\begin{split}
&\frac{1}{z^{2}}\frac{1}{\phi(a;1,\frac{u}{z})\phi(b;z,\frac{u}{z})}-\frac{1}{\phi(a;z,u)\phi(b;z,u)}\\
=&\frac{1}{u}\left\{\frac{a}{u}\frac{1}{\phi(ba^{-1};1,1)}\delta\left(\frac{u}{az}\right)+\frac{b}{u}\frac{1}{\phi(ab^{-1};1,1)}\delta\left(\frac{u}{bz}\right)\right\},
\end{split}
\end{align}
and
\begin{align}
\frac{1}{z}\frac{1}{\phi(a;z,u/z)}-\frac{1}{\phi(a;z,u)}=\frac{a^{\frac{1}{2}}}{u}\delta\left(\frac{u}{az}\right).
\end{align}

\section{3d crystal melting}\label{sec:3d_crystal}
In Section \ref{sec:QYreview}, quiver Yangian was defined by a quiver diagram and loop constraints.
The quiver diagram and loop constraints are constructed from a toric diagram \cite{Ooguri_2009,Li:2020rij}.
In this Appendix, we summarize this procedure\footnote{
The brane configuration of toric CY 3-folds and its relation with the quiver diagram was originally introduced in \cite{Franco:2005rj}.}. See \cite{Ooguri_2009} for details and physical interpretations.

\subsection{From toric diagram to quiver diagram}

\begin{figure}[h]
    \begin{tabular}{cc}
      \begin{minipage}{0.45\hsize}
        \centering
        \includegraphics[width=3cm]{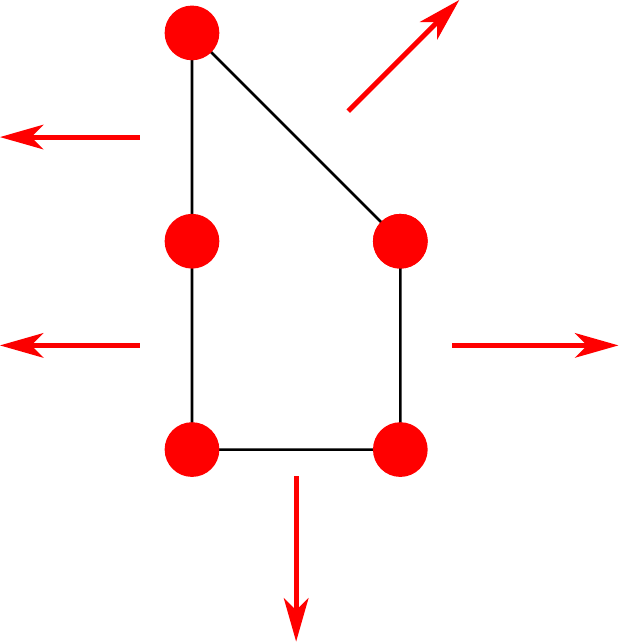}
        \subcaption{Toric diagram}
      \end{minipage} &
      \begin{minipage}{0.45\hsize}
        \centering
        \includegraphics[width=3.5cm]{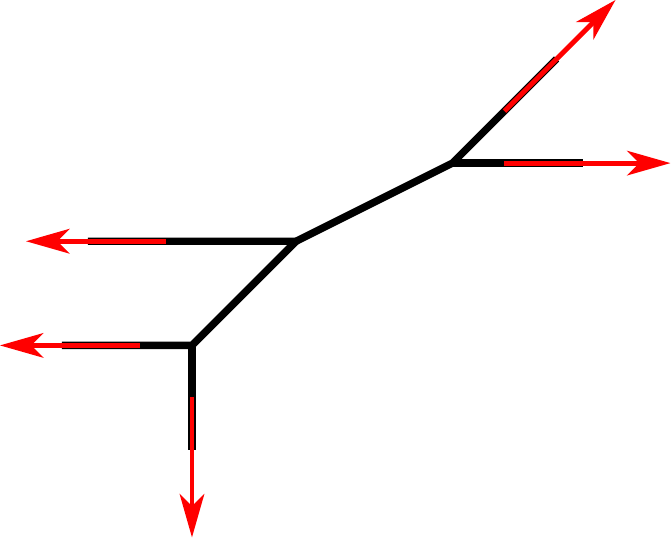}
        \subcaption{Web diagram}
      \end{minipage}\\
      \begin{minipage}{0.45\hsize}
        \centering
        \includegraphics[width=3cm]{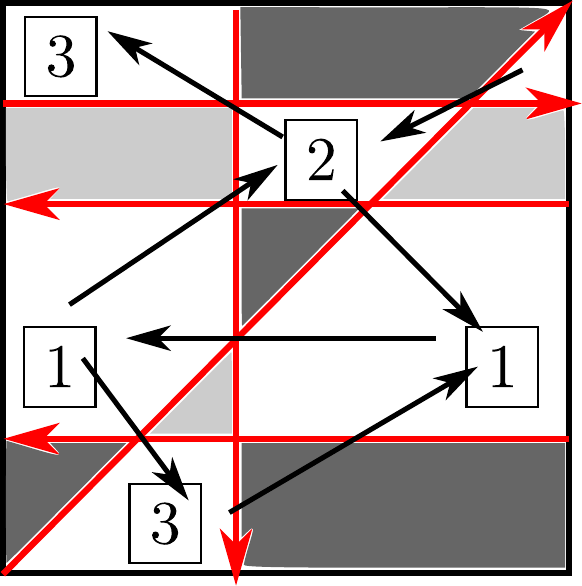}
        \subcaption{Brane configuration}
      \end{minipage} &
      \begin{minipage}{0.45\hsize}
        \centering
        \includegraphics[width=4cm]{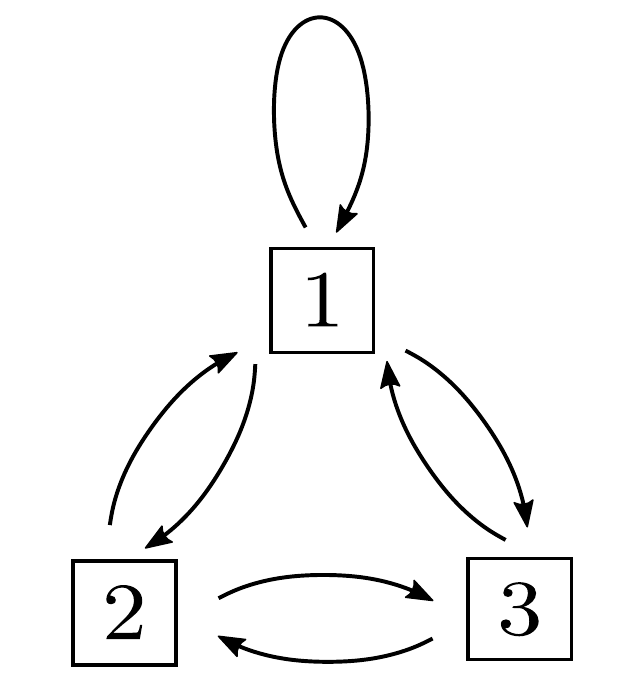}
        \subcaption{Quiver diagram}
      \end{minipage}
    \end{tabular}
\caption{Toric diagram, web diagram, brane configuration, and quiver diagram for the Suspended Pinch Point (SPP) singularity. (a) Toric diagram for the SPP. (b) The web diagram of the SPP. It is dual to the toric diagram. The red arrows are called external lines from now on. (c) The rectangle is a fundamental domain of the torus $\mathbb{T}^{2}$ and the boundary is identified periodically. The red lines physically correspond to NS5-branes, which divide the fundamental domain into several domains: dark gray, light gray, and white. The white regions are assigned vertices, and there are arrows between each vertex. The orientation of the arrows is defined so that the light gray domain is always on the left of the arrowhead. (d) Quiver diagram obtained from the brane configuration.}
\label{fig:gl21_3d}
\end{figure}
We start with a toric diagram and draw an outward red line perpendicular to each arrow of the toric diagram.
Each red lines express an NS5-brane, and the direction of the arrows means the direction of the NS5-brane.
These correspond to the external lines of the web diagram in the context of the topological string.

\paragraph{Brane configuration}
Now that we know how many NS5-branes we have and which direction they face, we consider their configuration on a torus $\mathbb{T}^2$.
First, we draw a periodic square region where the top and bottom, left and right, are identical, which expresses a torus.

Next, we place the red lines corresponding to the arrows of the toric diagram, which express NS5-branes.
Because of the constraint on NS5-charge, arbitrary configurations of NS5-branes are not allowed.
To describe the constraint, we paint each region white, dark gray, or light gray as an indicator of the NS5-branes orientation.
We paint dark gray if the red boundary lines are all counter-clockwise, light gray if all clockwise, and white otherwise.

The constraints on the placement of the red lines are as follows:
\begin{itemize}
    \item Two lines can intersect, but three or more lines must not intersect at a single point.
    \item White regions can connect by points, but not by lines.
\end{itemize}
If any of the above conditions are not satisfied, we need to change the configuration of the red lines to satisfy them.

\paragraph{Quiver diagram}
Taking the low energy limit in this brane configuration, we obtain the quiver gauge theory.
The quiver diagram characterizes the quiver gauge theory.

Each white region in the brane configuration corresponds to the vector multiplet in the quiver gauge theory, and each connection between the white regions corresponds to the chiral multiplet.
In other words, we obtain the quiver diagram by denoting the white regions as vertices and the connections between them as arrows.

\begin{figure}
    \centering
    \includegraphics[width=10cm]{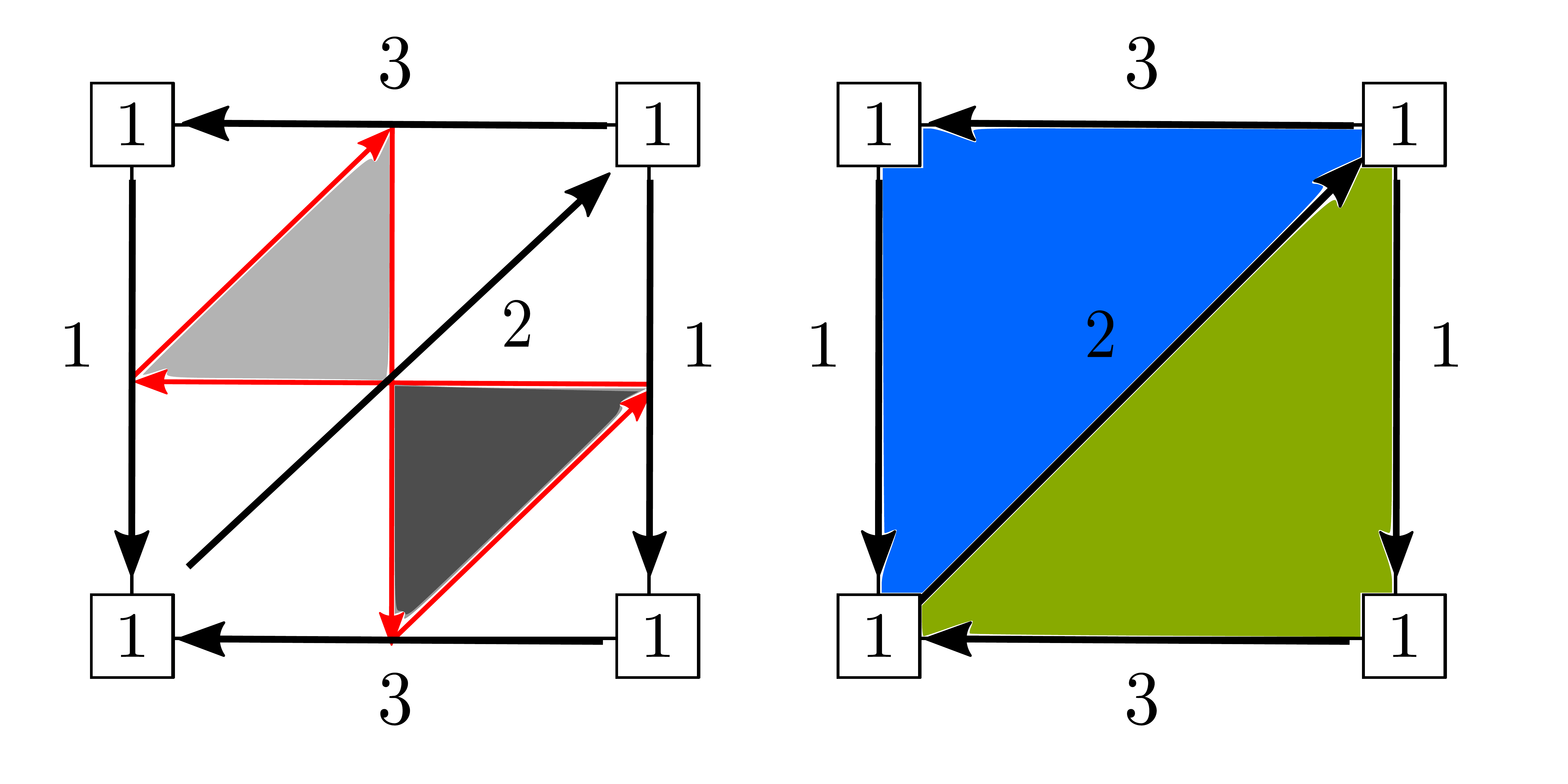}
    \caption{$Q_{0}$, $Q_{1}$, $Q_{2}$ of the $\mathbb{C}^{3}$ geometry. There is only one vertex here, $Q_{0}=\{1\}$. The set of arrows connecting vertices are $Q_1=\{1\xrightarrow{1} 1,1\xrightarrow{2} 1,1\xrightarrow{3} 1\}$. Since we are considering the torus, which means we are identifying the boundaries, the vertices and edges on the boundaries are identified. We have two elements of $Q_{2}$ and they are the blue and green region. They are identified with the closed loop of arrows $1\xrightarrow{1} 1 \xrightarrow{2} 1 \xrightarrow{3} 1$ and $1\xrightarrow{3} 1 \xrightarrow{2} 1 \xrightarrow{1} 1$ respectively. Thus, $Q_2=\{1\xrightarrow{1} 1 \xrightarrow{2} 1 \xrightarrow{3} 1,1\xrightarrow{3} 1 \xrightarrow{2} 1 \xrightarrow{1} 1\}$.}
    \label{fig:gl1Q2}
\end{figure}

We assign numbers $1, 2, \cdots, |Q_0|$ to each white region to distinguish the vertices.
We denote the set of these vertices by $Q_0$.
Next, we draw arrows between the regions of $Q_0$ connected by a single point (not by a line) on the brane configuration. The orientation of the arrows is chosen so that the dark (resp. light) gray region is always on the right (resp. left) of the arrowhead. We denote the set of these arrows by $Q_1$.
After writing down $Q_{0}$ and $Q_{1}$ on the torus, one will see that the two-dimensional surface of the torus is decomposed into areas surrounded by the arrows of $Q_{1}$. We denote the set of these areas as $Q_{2}$. The element of $Q_{2}$ is identified with the sequence of arrows surrounding it.
The quiver diagram $Q$ is the combination of $Q_0$, $Q_1$, and $Q_2$,\footnote{While the combination of $Q_0$ and $Q_1$ is usually called a quiver diagram, $Q_2$ is also obtained from a toric diagram, so this paper calls $(Q_0, Q_1, Q_2)$ a quiver diagram for convenience.}
\begin{equation}
    Q=(Q_0, Q_1, Q_2).
\end{equation}
We take the $\mathbb{C}^{3}$ geometry for an example (see Figure \ref{fig:gl1Q2}). In this case, we only have one vertex $Q_{0}=\{1\}$ and three arrows $Q_1=\{1\xrightarrow{1} 1,1\xrightarrow{2} 1,1\xrightarrow{3} 1\}$. The areas decomposed by the arrows are the blue and green region, and they are identified with the sequences of arrows $1\xrightarrow{1} 1 \xrightarrow{2} 1 \xrightarrow{3} 1$ and $1\xrightarrow{3} 1 \xrightarrow{2} 1 \xrightarrow{1} 1$ respectively.

\subsection{Loop constraints}
Each white region in the brane configuration corresponds to an atom in the 3d crystal.
Then, a path along with the arrows on the brane configuration corresponds to a route in the 3d crystal.
From only the quiver diagram, it is impossible to distinguish between the case where the route wraps around the torus non-trivially and the route loops around the same point.
We introduce loop constraints to supplement such information.

Each element of $Q_2$ defines a loop constraint. For the $\mathbb{C}^{3}$ geometry, the parameter $h_{11}^{(i)}$ is associated to the arrow $1 \xrightarrow{i} 1$ (see section \ref{sec:QYreview}), and there is only one independent loop constraint 
\begin{equation}
    h_{11}^{(1)}+h_{11}^{(2)}+h_{11}^{(3)}=0.
\end{equation}
One might think that we obtain two loop constraints from both the blue and green regions, but since we are considering the torus, after imposing one constraint on the blue (resp. green) region, the other constraint on the green (resp. blue) region will be automatically satisfied. 

One can do the same thing for Figure \ref{fig:gl21_3d} and obtain loop constraints (\ref{eq:loopconst_eq1})-(\ref{eq:loopconst_eq4}).

\section{KK relation and KE relation}\label{sec:appendix_bootstrapp}
\subsection{KK relation}
Since $K_{i}^{\pm}(z)$ acts diagonally on the crystal configuration $\Lambda$,
\begin{align}
    K_{i}^{\pm}(z)K_{j}^{\pm}(w)\ket{\Lambda}=K_{j}^{\pm}(w)K_{i}^{\pm}(z)\ket{\Lambda}
\end{align}
we obtain 
\begin{align}\label{eq:KKrelation}
    K_{i}^{\pm}(z)K_{j}^{\pm}(w)=K_{j}^{\pm}(w)K_{i}^{\pm}(z).
\end{align}
The property that $K_{i}^{\pm}(z)$ acts diagonally on the crystal configuration is one of the consequence of setting one of the central charges $C=1$ in (\ref{eq:defofQuiverAlgebra}). Since we are considering a representation with $C=1$, the KE relation of $K_{i}^{+}(z)$ and $K_{i}^{-}(z)$ are the same, and we express either of them as $K_{i}(z)$.
We also omit the sign $[\quad]_{\pm}$.
\subsection{KE and KF relation}
Apply first  $E_{j}(w)$ and then $K_{i}(z)$, then we get 
\begin{align}
    \begin{split}
        &K_{i}(z)E_{j}(w)\ket{\Lambda}\\
        =&\sum_{\fbox{$j$}\in \text{Add}(\Lambda)}\Psi_{\Lambda+\fbox{$j$}}^{(i)}(z,u)\epsilon(\Lambda\rightarrow\Lambda+\fbox{$j$})\sqrt{p^{(j)}\underset{x=uq(\fbox{$j$})}{\Res}\Psi_{\Lambda}^{(j)}(x,u)}\delta\left(\frac{w}{uq(\fbox{$j$})}\right)\ket{\Lambda+\fbox{$j$}}.
        \end{split}
        \end{align}
Applying the operators in a different order gives
\begin{align}
        \begin{split}
        &E_{j}(w)K_{i}(z)\ket{\Lambda}\\
        =&\sum_{\fbox{$j$}\in \text{Add}(\Lambda)}\Psi_{\Lambda}^{(i)}(z,u)\epsilon(\Lambda\rightarrow\Lambda+\fbox{$j$})\sqrt{p^{(j)}\underset{x=uq(\fbox{$j$})}{\Res}\Psi_{\Lambda}^{(j)}(x,u)}\delta\left(\frac{w}{uq(\fbox{$j$})}\right)\ket{\Lambda+\fbox{$j$}}
    \end{split}
\end{align}
Using (\ref{eq:chargeansatz}), we will see
\begin{align}
\begin{split}
\Psi_{\Lambda+\fbox{$j$}}^{(i)}(z,u)\delta\left(\frac{w}{uq(\fbox{$j$})}\right)&=\varphi^{j\Rightarrow i}(z,uq(\fbox{$j$}))\Psi_{\Lambda}^{(i)}(z,u)\delta\left(\frac{w}{uq(\fbox{$j$})}\right)\\
&=\varphi^{j\Rightarrow i}(z,w)\Psi_{\Lambda}^{(i)}(z,u)\delta\left(\frac{w}{uq(\fbox{$j$})}\right)
\end{split}
\end{align}
and obtain
\begin{align}\label{eq:KErelation}
    K_{i}(z)E_{j}(w)=\varphi^{j\Rightarrow i}(z,w)E_{j}(w)K_{i}(z).
\end{align}

Similarly we obtain the relation of $K_{i}(z)$ and $F_{j}(w)$:
\begin{align}\label{eq:KFrelation}
     K_{i}(z)F_{j}(w)=(\varphi^{j\Rightarrow i}(z,w))^{-1}F_{j}(w)K_{i}(z).
\end{align}

\bibliographystyle{JHEP}
\bibliography{QQTA}
\end{document}